\documentclass[fleqn,usenatbib]{mnras}

\usepackage{newtxtext,newtxmath}

\usepackage[T1]{fontenc}
\usepackage{ae,aecompl}

\usepackage{graphicx}	
\usepackage{amsmath}	
\usepackage{amssymb}	
\usepackage[version=4]{mhchem}  
\usepackage[utf8x]{inputenc}    

\title[Ion desorption induced by cosmic-rays bombardment]{Low-temperature chemistry induced by cosmic rays: positive and negative ions desorption from nitrile-bearing astrophysical ice analogues}

\author[F. de A. Ribeiro et al.]{
Fabio de A. Ribeiro,$^{1,2}$\thanks{E-mail: fabio.ribeiro@ifrj.edu.br (FAR)}
Guilherme C. Almeida,$^{3}$
Wania Wolff,$^{2}$
\newauthor
Heloisa M. Boechat-Roberty,$^{4}$
Maria L. M. Rocco$^{3}$
and Enio F. da Silveira$^{5}$\\
$^{1}$Instituto Federal de Educação, Ciência e Tecnologia do Rio de Janeiro, Rua Cel. Delio Menezes Porto, 1045, Nilópolis, RJ 26530-060, Brazil\\
$^{2}$Instituto de Física, Universidade Federal do Rio de Janeiro, Av. Athos da Silveira Ramos, 149, Rio de Janeiro, RJ 21941-909, Brasil\\
$^{3}$Instituto de Química, Universidade Federal do Rio de Janeiro, Av. Athos da Silveira Ramos, 149, Rio de Janeiro, RJ 21941-909, Brasil\\
$^{4}$Observatório do Valongo, Universidade Federal do Rio de Janeiro, Ledeira Pedro Antonio, 43, Rio de Janeiro, RJ 20080-090, Brasil\\
$^{5}$Departamento de Física, Pontifícia Universidade Católica do Rio de Janeiro, Rua Marquês de São Vicente, 225, Rio de Janeiro, RJ 22453-900, Brazil\\
}

\date{Accepted XXX. Received YYY; in original form ZZZ}

\pubyear{2019}

\begin{document}
\label{firstpage}
\pagerange{\pageref{firstpage}--\pageref{lastpage}}
\maketitle

\begin{abstract}
In cold core of dark molecular clouds, where the UV radiation from external sources is strongly attenuated, cosmic rays can induce chemical reactions on the surface of ice-covered grains promoting the ejection of the processed material to the gas phase. We report the positive and negative secondary ion emission from pure CH$_3$CN, C$_2$H$_3$CN and i−C$_3$H$_7$CN ices due to the bombardment of heavy ions ($^{252}$Cf fission fragments), simulating the incidence of cosmic rays onto icy surfaces. The secondary ions emitted from each sample were analysed by time-of-flight mass spectrometry (TOF-MS), using Plasma Desorption Mass Spectrometry (PDMS) technique. Several ionic species were identified, indicating strong fragmentation on the frozen surface. Proton-transfer processes are suggested to play a role for positive ion desorption, as evidenced by the protonated RCNH$^{+}$ parent molecules and (RCN)$_n$H$^+$ ionic clusters. The high electron affinity of the cyano radical seems to contribute to the strong emission of CN$^-$, as well as anions attributed to the CH$_m$CN$^-$ fragment and (RCN)$_n$CN$^-$ cluster series. Sputtering and desorption of ion clusters (positive and negative) induced by heavy ion bombardment are suggested to constitute a route by which new neutral or ionised molecular species may be delivered to the gas phase where thermal desorption is negligible.
\end{abstract}

\begin{keywords}
astrochemistry -- methods: laboratory: solid state --  molecular processes -- (ISM:) cosmic rays 
\end{keywords}



\section{Introduction}

Nitriles, also known as cyanides, are known to be present in a variety of sources in the interstellar medium (ISM) and in circumstellar environments. Among the many astrophysical relevant nitriles, acetonitrile (CH$_3$CN) is the simplest organic molecule bearing the nitrile or cyano (C≡N) functional group. CH$_3$CN was first detected towards Sgr A and Sgr B molecular clouds \citep{Solomon1971}, then it was also detected toward hot cores and corinos, associated with high \citep[e.g.][]{Pankonin2001,Araya2005, Rosero2013} and low-mass \citep[e.g.][]{Cazaux2003,Taquet2015} star-formation, including their prototype, the Orion hot core \citep{Bell2014}. Acrylonitrile (C$_2$H$_3$CN), which contains an unsaturated C=C bond in its molecular structure, is also present in hot cores, such as those within the Orion KL and W51 North massive star-forming regions, where its abundance is correlated to the closely related saturated nitrile, propionitrile (C$_2$H$_5$CN) \citep[e.g.][]{Fontani2007,Rong2015}. Butyronitrile (C$_3$H$_7$CN) may exist as the normal straight-chain (n-C$_3$H$_7$CN) or the branched isobutyronitrile (i-C$_3$H$_7$CN) molecule. i-C$_3$H$_7$CN was detected towards the Sgr B2(N) molecular cloud \citep{Belloche2014}, constituting the first branched molecule to be discovered in the ISM. Even though the linear isomer n-C$_3$H$_7$CN was previously known towards Sgr B2(N) \citep{Belloche2009}, both isomers were detected in the Orion hot core \citep{Pagani2017}. However, despite similar molecular abundance ratios C$_2$H$_3$CN/CH$_3$CN and C$_2$H$_5$CN/CH$_3$CN are observed toward Sgr B2 and Orion KL star-forming regions, a variation by a factor of 2 was found when compared to the same ratio measured for the quiescent giant molecular cloud G+0.693 \citep{Zeng2018}. This reinforces that though nitriles are ubiquitous towards many line-of-sights, the mechanisms ruling their formation and processing may differ considerably depending on the physical conditions of the source.\\
Gas phase chemistry alone is unable to account for the high abundances of many of the complex organic molecules (COMs) observed, especially in the case of highly saturated molecules, such as the aliphatic nitriles CH$_3$CN and C$_2$H$_5$CN \citep{Bisschop2007}. This has led to suggest that many of the COMs seen in hot cores are synthesised mainly by solid state reactions on ice mantles covering dust grains, and then released to the gas phase upon the warming up of the protostellar envelope \citep{Garrod2008}. The alternative formation pathway involves the chemical synthesis on the surface of grains by successive recombination and hydrogenation of frozen CN, C$_2$N as well as C$_2$H$_2$, C$_2$H$_4$ and HC$_3$N radicals on mantles followed by desorption to the gas phase \citep{Garrod2013,Garrod2017}. However, the molecular inventory  will remain frozen until the temperature is high enough to promote the evaporation of ice mantles from grains. Indeed, high angular resolution observations of the emission of O- and N-bearing (including CH$_3$CN, C$_2$H$_3$CN and C$_2$H$_5$CN)  molecules in the disk around the high-mass protostar AFGL 4176 made by \cite{Bogelund2019} indicate that the inventory of complex molecules may already be set in the cold cloud stage. Besides warm environments such as hot cores, CH$_3$CN \citep{Matthews1983a}, C$_2$H$_3$CN \citep{Matthews1983b} and benzonitrile (c-C$_6$H$_5$CN) \citep{McGuire2018} were detected towards the TMC-1 dark cloud, where the gas and grain can reach temperatures as low as 10 K. Their presence in cold dark clouds such as the TMC-1 remains a puzzle, since the temperature is too low to induce thermal desorption or evaporation of ices. Also, theoretical models under-predict the abundance of nitriles by gas phase reactions involving only neutrals \citep{Vigren2009}, so that ion-molecule reactions are frequently evoked in order to account for the abundance of highly reactive species, including ions, radicals, and unsaturated species such as cyanopolyynes HC$_n$N in cold cores \citep{Zeng2018}.\\
Because of the high proton affinities of nitriles, protonated nitriles (RCNH$^+$) are expected to be promptly produced in the ISM. Even though the attempts to detect CH$_3$CNH$^+$ towards the ISM have failed \citep{Turner1990}, gas phase CH$_3$CNH$^+$ would contribute to maintain the high abundances observed for CH$_3$CN by dissociative electron recombination of CH$_3$CNH$^+$ \citep{Vigren2008}. Thus, dissociative recombination of the protonated nitriles RCNH$^+$ in the gas phase might contribute for producing the respective neutral nitriles (RCN). Linear carbon-chain anions C$_{2n+2}$H$^-$ and C$_{2n-1}$N$^-$ (n = 1, 2, 3) have been observed in the circumstellar envelope of IRC+10216 and might also be present in the TMC-1 dark cloud \citep{Thaddeus2008}, where the related neutrals C$_3$N and C$_4$H are known to be present. The inclusion of anions in the chemical modelling of dark clouds resulted on the enhancement of the abundances of carbon-chain molecules, C$_n$, C$_n$H, C$_n$H$_2$, C$_n$N, and HC$_n$N, by providing new formation routes to the larger members of each family via smaller molecules \citep{Catherine2009}. Though no surface chemistry was added, \cite{Catherine2009} obtained a better agreement of cyanopolyyne abundances observed in TMC-1 by including anionic species.\\
Experiments regarding ion-neutral reactions \citep{Petrie1992}, dissociative recombination on gas phase \citep{Vigren2009} and the energetic processing of nitriles within ice matrices \citep{Hudson2004,Moore2010,Toumi2014,Couturier-Tamburelli2015,Toumi2016} are important to constrain physical parameters in chemical models and retrieve optical constants for astronomical observations. Experiments concerning the emission of ions from the surface of ices to the gas can complement such models. Ion desorption rates might be of valuable importance as an input for ion-neutral reactions modelling the abundance of nitriles and related compounds in cold environments.\\
In cold cores such as TMC-1, chemical reactions on the surface of ice mantles are suppressed by the little thermal energy that limit atomic and molecular diffusion. On the other hand, while the UV radiation from external sources is strongly attenuated, cosmic rays (MeV - TeV) can influence the chemistry of the gas and solid components of the interstellar matter, as they can deeply penetrate into the cloud \citep{Gaches2019}.\\
One of the physico-chemical effects caused by the incidence of cosmic rays onto dust grains is the ejection of atoms and molecules from the ice mantle to the gas phase \citep{Herbst2006,Collings2012,Iqbal2018}. Charged particles (projectiles) moving through a solid collide randomly with the target species along their track through Coulomb interactions with the nuclei and electrons of the target. The projectile energy loss per unit path length is the stopping power, whose magnitude depends both on the projectile and the medium. It is customary to assume that the energy transferred by the ion to the target can be separated into contributions due to elastic nuclear collisions (momentum transfer to the target nuclei) and inelastic collisions (electronic excitation of the target electrons), the so-called nuclear and electronic stopping power, respectively \citep{Nastasi1996,Bringa2003}.
Depending on the energy relaxation processes within the solid after the ion-target interaction, the excitation energy released to the solid lattice generate repulsive states that may lead to extra kinetic energy to initiate sputtering and surface modification \citep{Bringa2003}. For charged particles with velocity larger than the Bohr velocity (v = 2.19 $\times$ 10$^{8}$ cm s$^{-1}$), the electronic stopping power dominates, and both energy and momentum from the projectile are transferred to target molecules by ion–electron interactions. The energy deposited in the region around the line of passage of the projectile leads to molecular electronic excitation, ionisation and emission of secondary electrons ($\delta$-electrons). Either the $\delta$-electrons generated or the repulsion between the unscreened nuclear charges may convert electronic energy into lattice motion \citep{Sundqvist1993,Bringa2003,Iza2006} that lead to the emission of neutrals and ions from the solid.\\
The energy distribution for cosmic-ray protons peaks around 10$^{8}$ – 10$^{9}$ eV \citep{Rab2017} and, thus, the electronic stopping power dominates for ion-surface astrophysical applications. Therefore, electronic sputtering results obtained experimentally might aid in the analysis of the effects induced by cosmic rays bombardment onto interstellar ices. Secondary ion emission probes the atomic and molecular species emitted just after the projectile impact (about $\thicksim$10$^{-12}$ s) \citep{Sundqvist1993}. Though the majority of particles are ejected as neutrals \citep{Rothard2017}, many laboratory works have reported the role of electronic sputtering in the release of more complex molecules, such as ionic molecular clusters \citep[e.g.][]{Farenzena2005b,Ponciano2006,Andrade2007,Andrade2009, Martinez2014} and PAHs in their neutral and cationic forms and cationic fullerenes \citep{Pino2019}.
In this work, high-mass ions, resulting from $^{252}$Cf nuclear fission, are employed to bombard and sputter secondary ions from ice analogues constituted by CH$_3$CN, C$_2$H$_3$CN and i-C$_3$H$_7$CN, respectively, on the basis of the Plasma Desorption Mass Spectrometry (PDMS) technique. The influence of the alkyl and vinyl substituent on the desorption of positive and negative secondary ions is discussed.

\section{Experimental Methods}

\begin{figure}
	\includegraphics[width=\columnwidth]{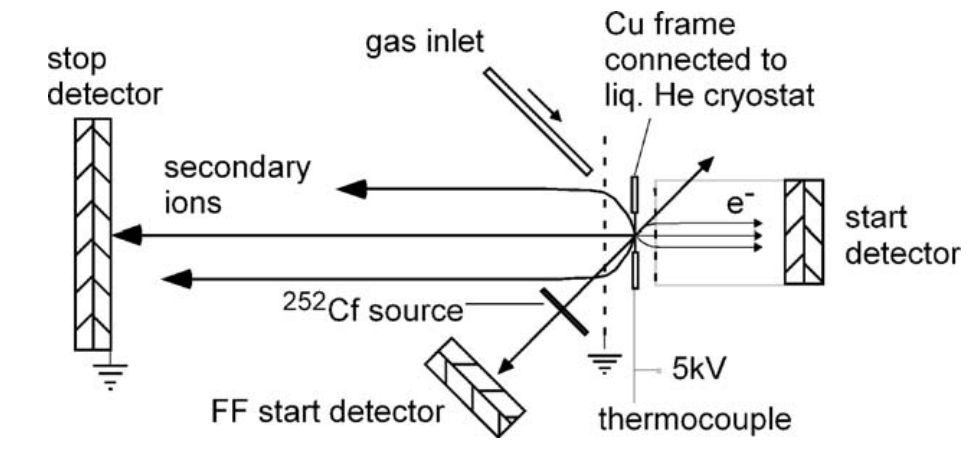}
    \caption{Scheme of the experimental setup employed to analyse secondary ion desorption by means of the PDMS technique. Targets were frozen at 60 K and bombarded by fission fragments resulting from the $^{252}$Cf nuclear decay ($^{252}$Cf-FFs). Adapted from \citet{Farenzena2006}.}
    \label{Fig.1}
\end{figure}

Secondary ion desorption induced by heavy ion bombardment was studied by the Plasma Desorption Mass Spectrometry (PDMS) technique at the Van de Graaff Laboratory of the Pontifícia Universidade Católica (PUC-Rio), at Rio de Janeiro, Brazil. The main parts of the experimental setup consist of a $^{252}$Cf ionisation source, a time-of-flight (TOF) mass spectrometer, and a closed-cycle He cryostat with a copper frame connected to the cold finger (Fig.~\ref{Fig.1}). All instruments are housed in high vacuum conditions under a pressure of better than 10$^{-7}$ mbar. More details about the instrumentation can be found elsewhere \citep{Collado2004,Farenzena2006}.\\

The $^{252}$Cf source has a half-life time of 2.65 years. It decays mainly by the emission of 6 MeV alpha particles (97\%) or by two simultaneous (3\%) multiply charged fission fragments (hereafter $^{252}$Cf–FFs), emitted in opposite directions. The mean atomic weight, <\textit{A}>, of the light and heavy $^{252}$Cf–FFs peaks at <\textit{A}> = 106.5 u and 141.8 u, respectively, while the mean nuclear charge, <\textit{Z}>, peaks at <\textit{Z}> = 42.7 and 55.4, respectively. Typical $^{252}$Cf–FFs are $^{106}$Tc$^{22+}$ and $^{142}$Ba$^{18+}$, with energies of about 100 MeV and 80 MeV, respectively \citep{Macfarlane1976,Sundqvist1985,Knyazheva2006}. After traversing a thin  nickel foil, they impinge onto the frozen sample and induce the emission of neutrals and secondary ions. The secondary ions were extracted from the frozen target and directed toward the TOF spectrometer by applying a +5 kV electrical potential for the cations, while the anions were extracted by reversing the electrical potential to –5 kV.\\
The start signal for the time-of-flight of secondary ions was given by the electrons emitted at the backside of the copper frame, driving them onto a micro-channel plate detector. For the detection of secondary positive ions, this was achieved by polarising the electron grid and the start detector by +6.2 kV. For the detection of secondary negative ions, the polarisation of the electron grid and the start detector was set to –4 kV. In both cases, only those ions that reach the stop detector in coincidence with the electrons extracted from the back end of the substrate generate a TOF peak and are correctly counted (see Fig.~\ref{Fig.1}).\\
The ice analogues were grown in situ by condensation of the sample vapour onto a cooled copper frame, whose thickness is about 0.1 μm, directly mounted on the He-cryostat. The substrate temperature was set to 60 K to avoid N$_2$ condensation. Contamination by background gases was reduced by keeping a steady sample vapour streaming onto the substrate, thus continuously refreshing the ice surface during the experiments.\\
To report the ion desorption yield as ions emitted per fission fragment impact, we measured the PDMS mass spectra of the ices formed by the condensation of the residual gas on the copper substrate at 60 K, and then compared to the PDMS mass spectra reported by \cite{Collado2004}. A good agreement between the present relative intensities and that of the {(H$_2$O)$_n$H$^+$} (n = 1–18) ion series was observed, revealing the condensation of water vapour from the residual gas. The relative ion desorption yield values for the {(H$_2$O)$_n$H$^+$} ion series were normalised to the results reported by \cite{Collado2004}. The derived scaling factor was used to express the desorption yield values of the present nitriles data as secondary ions per heavy ion impact.\\

\section{Results and Discussion}
\subsection{Desorption of positive ions}

\begin{figure}
	\includegraphics[width=\columnwidth]{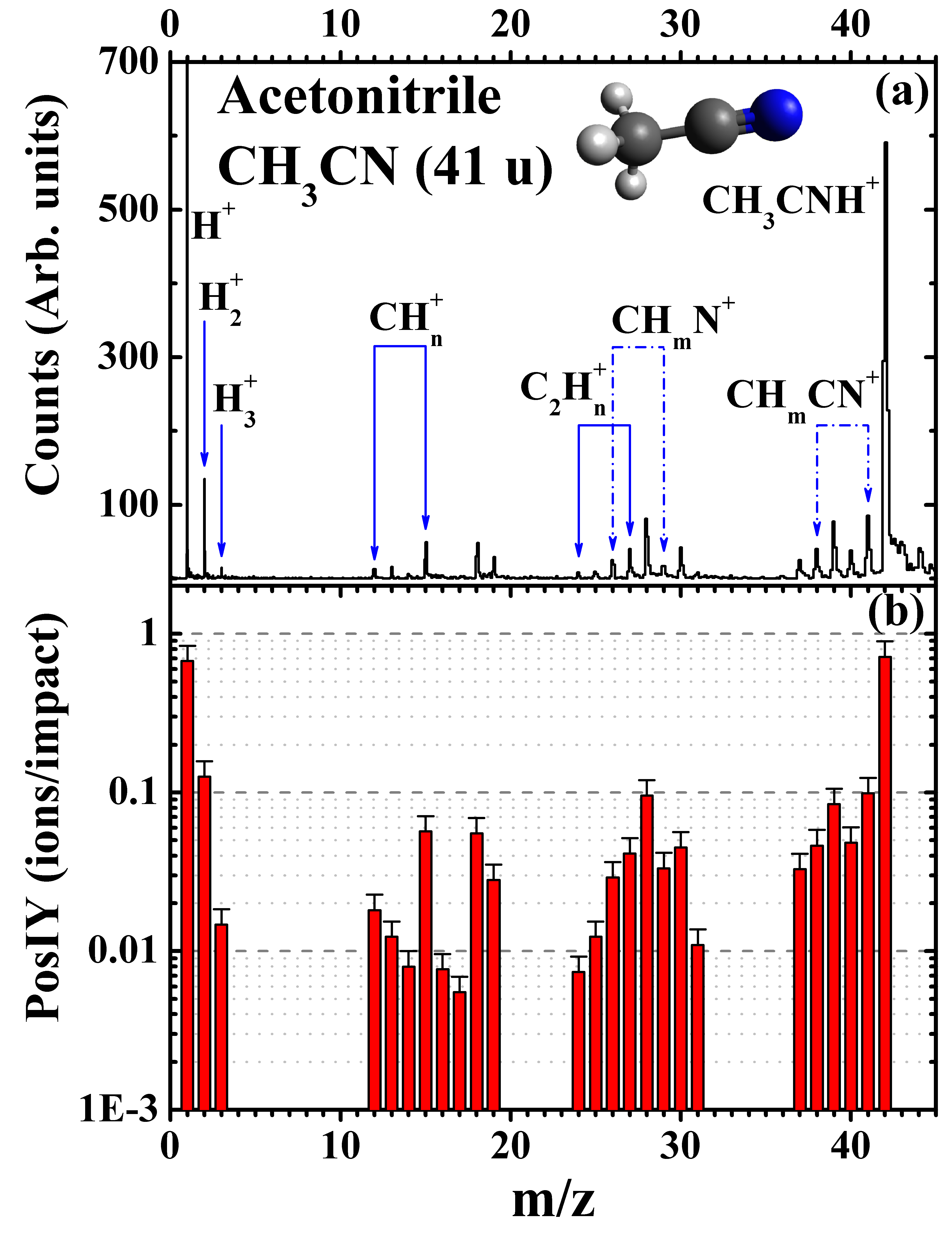}
    \caption{Top panel(a): Mass spectrum of positive secondary ions desorbed from the acetonitrile (CH$_3$CN) ice at 60 K. Besides the H$_n^+$ (n = 1-3) ions series, one can note the CH$_n^+$ (n = 0-3), C$_2$H$_n^+$ (n = 0-3)/CH$_m$N$^+$ (n = 0-3), C$_2$H$_mN^+$ (n = 0-3) and the protonated parent ion CH$_3$CNH$^+$. Bottom panel (b): Positive ion yield (PosIY, ions/impact) for each desorbed fragment.}
    \label{Fig.2}
\end{figure}

\begin{figure}
	\includegraphics[width=\columnwidth]{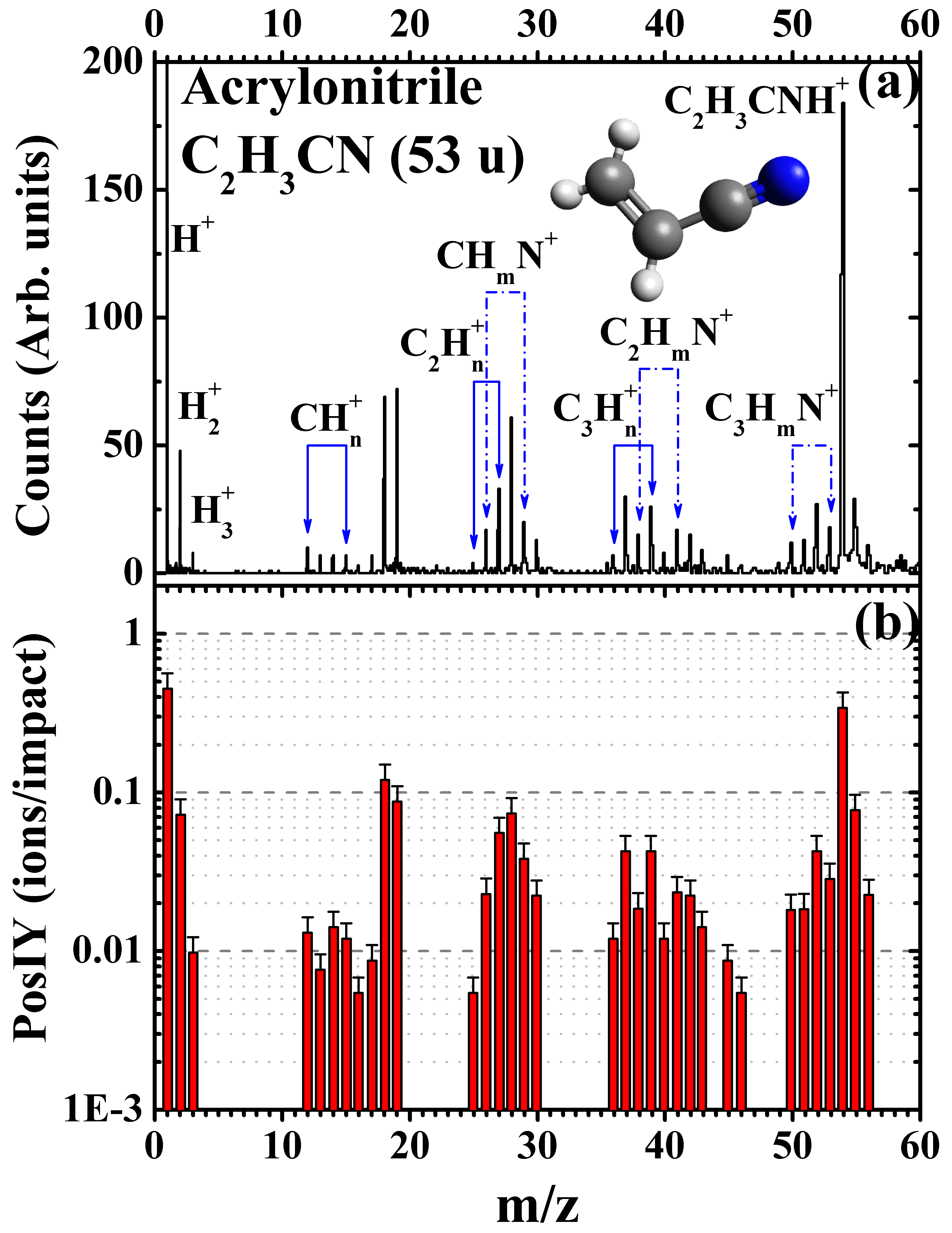}
    \caption{Top panel(a): Mass spectrum of positive secondary ions desorbed from acrylonitrile (C$_2$H$_3$CN) ice at 60 K. Besides the H$_n^+$ (n = 0-3) ions series, one can note the CH$_n^+$ (n = 0-3), C$_2$H$_n^+$ (n = 1-3)/CH$_m$N$^+$ (n = 0-3), C$_3$H$_n^+$ (n = 0-3)/C$_2$H$_m$N$^+$ (n = 0-3), C$_3$H$_m$N$^+$ (n = 0-3) and the protonated parent ion C$_2$H$_3$CNH$^+$. Bottom panel (b): Positive ion yield (PosIY, ions/impact) for each desorbed fragment.}
    \label{Fig.3}
\end{figure}

\begin{figure}
	\includegraphics[width=\columnwidth]{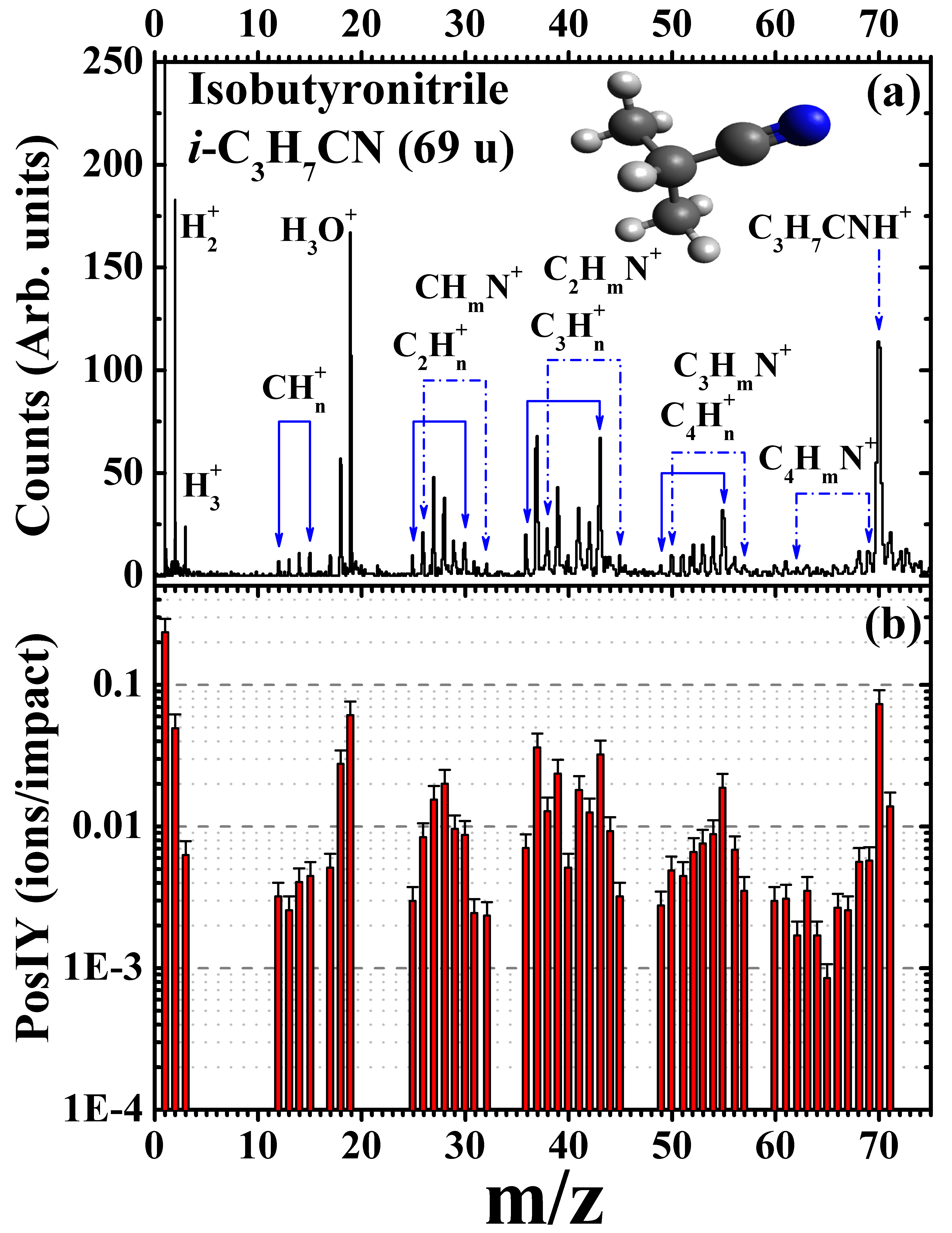}
    \caption{Top panel(a): Mass spectrum of positive secondary ions desorbed from isobutyronitrile (i-C$_3$H$_7$CN) ice at 60~K. Besides the $H_n^+$ (n = 0-3) ions series, one can note the CH$_n^+$ (n = 0-3), C$_2$H$_n^+$ (n = 1-6)/CH$_m$N$^+$ (n = 0-6), C$_3$H$_n^+$ (n = 0-7)/C$_2$H$_m$N$^+$ (n = 0-7), C$_4$H$_n^+$ (n = 1-7)/C$_3$H$_m$N$^+$ (n = 0-7), and the protonated parent ion C$_3$H$_7$CNH$^+$. Bottom panel (b): Positive ion yield (PosIY, ions/impact) for each desorbed fragment.}
    \label{Fig.4}
\end{figure}

The distributions of positive secondary ions from acetonitrile (CH$_3$CN), acrylonitrile (C$_2$H$_3$CN) and isobutyronitrile (\textit{i}-C$_3$H$_7$CN) ices emitted by bombardment with $^{252}$Cf fission fragments are shown in Fig.~\ref{Fig.2}a, Fig.~\ref{Fig.3}a and Fig.~\ref{Fig.4}a, respectively. The time-of-flight mass spectrum obtained is reported in the mass-to-charge ratio (m/z) scale up to the parent ion region. In each figure, the lower part (b) presents the positive ion yield (PosIY) derived for each desorbed fragment. The relative error for the PosIYs add up to ±25\% \citep{Farenzena2005a}.\\

\begin{figure}
	\includegraphics[width=\columnwidth]{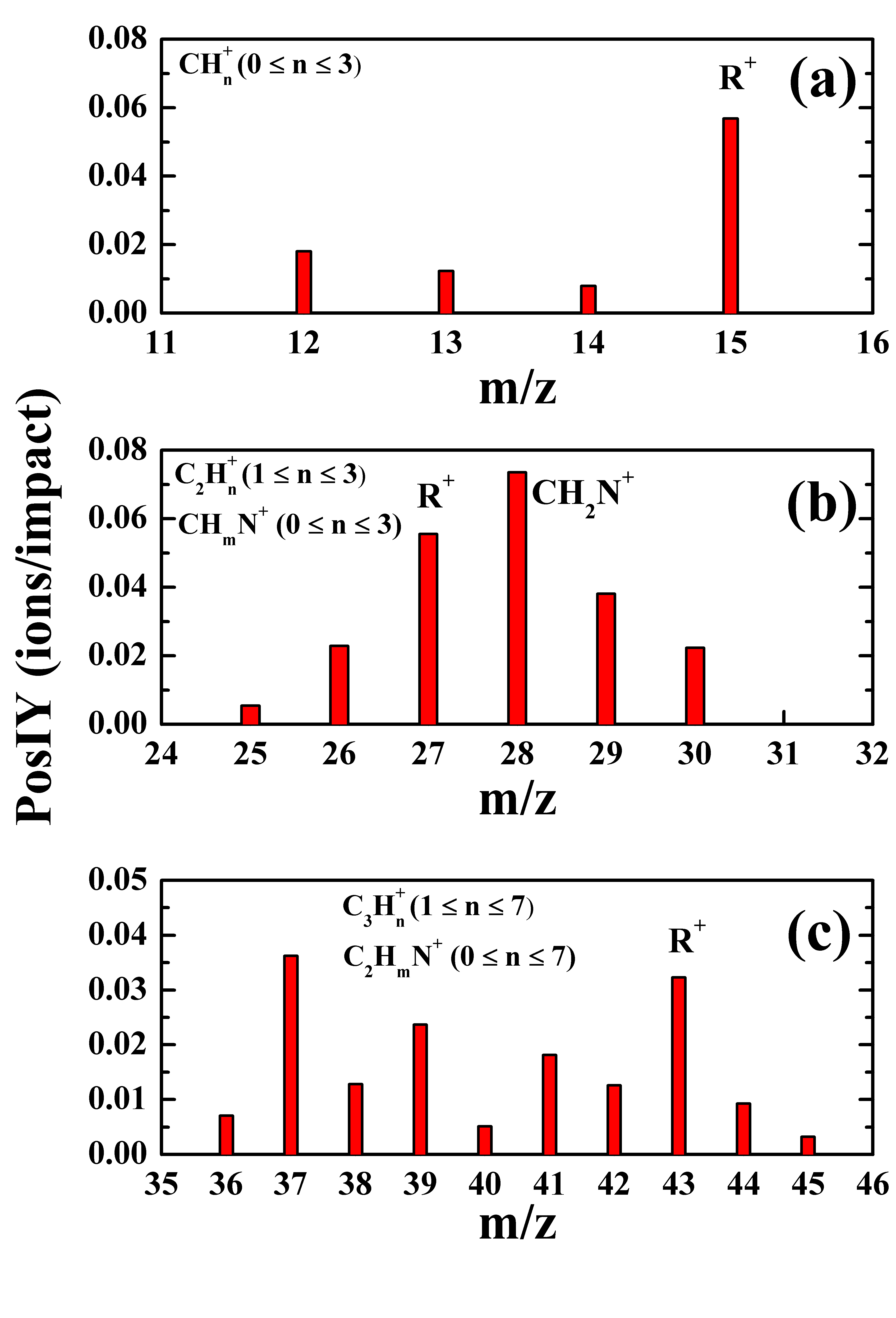}
    \caption{Positive ion yield (PosIY) distribution for the ions resulting from the loss of two atoms from the main chain of (a) acetonitrile, (b) acrylonitrile and (c) isobutyronitrile. R$^+$ corresponds to the CH$_3^+$ (m/z = 15), C$_2$H$_3^+$ (m/z = 27) and C$_3$H$_7^+$ (m/z = 43) ions, respectively.}
    \label{Fig.5}
\end{figure}

Aliphatic nitriles have the C$_x$H$_y$CN molecular structure, so each spectrum from Fig.~\ref{Fig.2}a to Fig.~\ref{Fig.4}a, can be divided into different ion series, according to the molecule stoichiometry. For the saturated aliphatic nitriles, such as acetonitrile and isobutyronitrile, $x$=1 and 3, respectively, and the maximum number of hydrogen per molecule is given by $y$ = 2$x$+1. For acrylonitrile, an unsaturated aliphatic nitrile, $x$ = 2 and $y$ = 2$x$–1. One can relate the number of ionic groups in the mass spectrum to the number of carbon and nitrogen atoms in each molecule, whereas the population of each ion series underlies on the processes leading to the fragmentation dynamics. The general fragmentation pattern for these molecules includes the H$_n^+$, C$_x$H$_n^+$ and C$_{x-1}$H$_m$N$^+$ ion series, in which each value of x defines a series, and n or m extends up to the value of y. However, the ions assigned as C$_x$H$_n^+$ and C$_{x-1}$H$_{n-2}$N$^+$ (2 $\leq$ n $\leq$ 2$x$+2) cannot be distinguished, since they cannot be separated with the present mass spectrometer.\\
The rupture of the R–CN$^+$ bond (where R is the alkyl or vinyl chain) and the loss of a CN radical is a predominant feature for the three nitriles. This may be inferred by the PosIY for the ions at $m/z$ = 15 (CH$_3^+$), $m/z$ = 27 (C$_2$H$_3^+$) and $m/z$ = 43 (C$_3$H$_7^+$) for CH$_3$CN, C$_2$H$_3$CN and i-C$_3$H$_7$CN, respectively, as shown in Fig.~\ref{Fig.5}. For CH$_3$CN (Fig.~\ref{Fig.5}a), the CH$_3^+$ has the largest PosIY among the CH$_n^+$ (n = 0-3) ionic distribution. Its desorption yield might also be related to the lower energy required to produce CH$_3^+$ (9.84 eV) \citep{LiasNIST} in respect to CN$^+$ (13.956 eV) \citep{Gans2017}. In the case of C$_2$H$_3$CN or i-C$_3$H$_7$CN, though the distribution does not peak exactly at C$_2$H$_3^+$ and C$_3$H$_7^{+}$ (R$^+$ in Fig.~\ref{Fig.5}b and Fig.~\ref{Fig.5}c, respectively), these two ions exhibit a considerable contribution within their respective groups, which contain the C$_x$H$_n^+$ and C$_{x-1}$H$_m$N$^+$ ion series. The loss of the CN radical and desorption of C$_3$H$_n^+$ ions from i-C$_3$H$_7$CN can also be inferred by the even-odd m/z intensity alternation in the interval 36 $\leq$ m/z $\leq$ 45 (Fig.~\ref{Fig.5}c). Such pattern is characteristic of the C$_x$H$_n^+$ ion series, since odd-numbered m/z hydrocarbon ions possess closed-shell electronic configuration, thus being more stable \citep{Betts1995}.\\

However, the strong contribution of m/z = 28 in the C$_2$H$_n^+$/CH$_m$N$^+$ ion series in Fig.~\ref{Fig.5}b indicates an efficient fragmentation channel on surface for desorption of the not-preformed CH$_2$N$^+$ fragment. Even though the peak at m/z = 28 may also be ascribed to either N$_2^{+}$ or CO$^{+}$ ion desorption, the temperature required for condensing N$_2$ or CO from the residual gas chamber would be considerably lower than the ice temperature under the present experimental conditions (60 K), since these volatile species desorb at temperatures < 30 K \citep{Collings2004,Burke2010,Martin-Domenech2014}. Another possible source for CO$^{+}$ ion desorption might arise from CO$_2$ co-deposition and fragmentation, but the mass spectrum of positive secondary ions desorbed from CO$_2$ by $^{252}$Cf-FFs is clearly absent \citep{Farenzena2006,Ponciano2006}. Therefore, even if entrapped, N$_2$, CO and CO$_2$ are expected to be very diluted within the ice matrix, and their contribution to the m/z = 28 peak is negligible. Thus, the CH$_2$N$^+$ ion may be produced by hydrogen migration onto the CN radical and loss of C$_2$H from C$_2$H$_3$CN$^+$. Desorption of CH$_2$N$^+$ also dominates the C$_2$H$_n^+$/CH$_m$N$^+$ ion series in CH$_3$CN and i-C$_3$H$_7$CN. This might be rationalised by the loss of CH and C$_3$H$_5$ radicals from CH$_3$CN$^+$ and C$_3$H$_7$CN$^+$, respectively. Considering the high energy involved in the dissociation of the cyano radical of 7.77 eV \citep{Costes1990}, the fragmentation of the alkyl and vinyl substituted nitriles seems to occur mainly by the process RCN$^+$ $\longrightarrow$ R$^+$ + CN or RCN$^+$ $\longrightarrow$ R + CN$^+$, depending on the charge localisation after molecular dissociation, thus preserving the CN bond.\\
The incidence of high energy $^{252}$Cf-FFs also induces the desorption of fragments that are not pre-formed on the isolated molecules. As can be seen in Fig.~\ref{Fig.2}a to Fig.~\ref{Fig.4}a, each ionic group extends up to higher mass, increasing the distribution of ions within the C$_x$H$_m$N$^+$ ion series. This includes, for example, the ions at m/z = 32 and m/z = 45 in the i-C$_3$H$_7$CN PDMS mass spectrum (Fig.~\ref{Fig.4}a). The former is attributed to the methylammonium cation (CH$_3$CNH$_3^+$), and the latter to C$_2$H$_7$N$^+$, which might be the ethylamine cation (C$_2$H$_5$NH$_2^+$). Thus, hydrogenation induced by heavy ion bombardment may lead to the desorption of amines from condensed nitriles.\\
As previously observed by electron stimulated ion desorption (ESID) of CH$_3$CN \citep{FAR2015}, the high intensity peak centred at m/z = 42 in Fig.~\ref{Fig.2}a is ascribed to the desorption of protonated acetonitrile, CH$_3$CNH$^+$. Through the later was observed in both ESID and PDMS, its relative intensity is different among these two techniques. The yield ratio CH$_3$CNH$^+$/ CH$_3$CN$^+$ in PDMS is almost five times that of ESID \citep{FAR2015}, showing stronger protonation during ion bombardment. As a general trend, C$_2$H$_3$CN and i-C$_3$H$_7$CN also desorb as protonated nitriles, RCNH$^+$. The yield ratio for desorption of the protonated and the single charged parent nitriles Y(RCNH$^+$)/Y(RCN$^+$) and the proton affinities evaluated for CH$_3$CN, C$_2$H$_3$CN and i-$C_3H_7CN$ compiled by \citep{Edward1998} are shown in Table~\ref{table1}.\\

\begin{table}
	\centering
	\caption{Observed ratio for the positive ion desorption yield of protonated, PosIY(RCNH$^+$), and single charged nitriles, PosIY(RCN$^+$) in $^{252}$Cf-PDMS for CH$_3$CN, C$_2$H$_3$CN and i-C$_3$H$_7$CN. Proton affinities were extracted from \citep{Edward1998}.}
	\label{table1}
	\begin{tabular}{lccr} 
		\hline
		Nitrile & ratio & Proton affinity (kJ~mol$^{-1}$)\\
		\hline
	CH$_3$CN & 7.26 & 779.2\\
		C$_2$H$_3$CN & 12.0 & 784.7\\
		i-C$_3$H$_7$CN & 12.8 & 803.6\\
		\hline
	\end{tabular}
\end{table}

Though the proton affinities do not vary to a large extent, they are sensitive to the aliphatic chain in the substituted nitriles. Noteworthy, the ratio Y(RCNH$^+$)/Y(RCN$^+$) increases as the substituent changes from methyl to isopropyl. The dissociation of the protonated parent molecule into smaller fragments is not discarded, because the fragments of the C$_x$H$_m$N$^+$ ion series with m greater than the number of hydrogen atoms in the parent molecule were observed (for instance, the peak m/z = 30, assigned to CH$_4$N$^+$, is present in the mass spectra of CH$_3$CN and C$_2$H$_3$N).\\

\subsection{Desorption of negative ions}

\begin{figure}
	\includegraphics[width=\columnwidth]{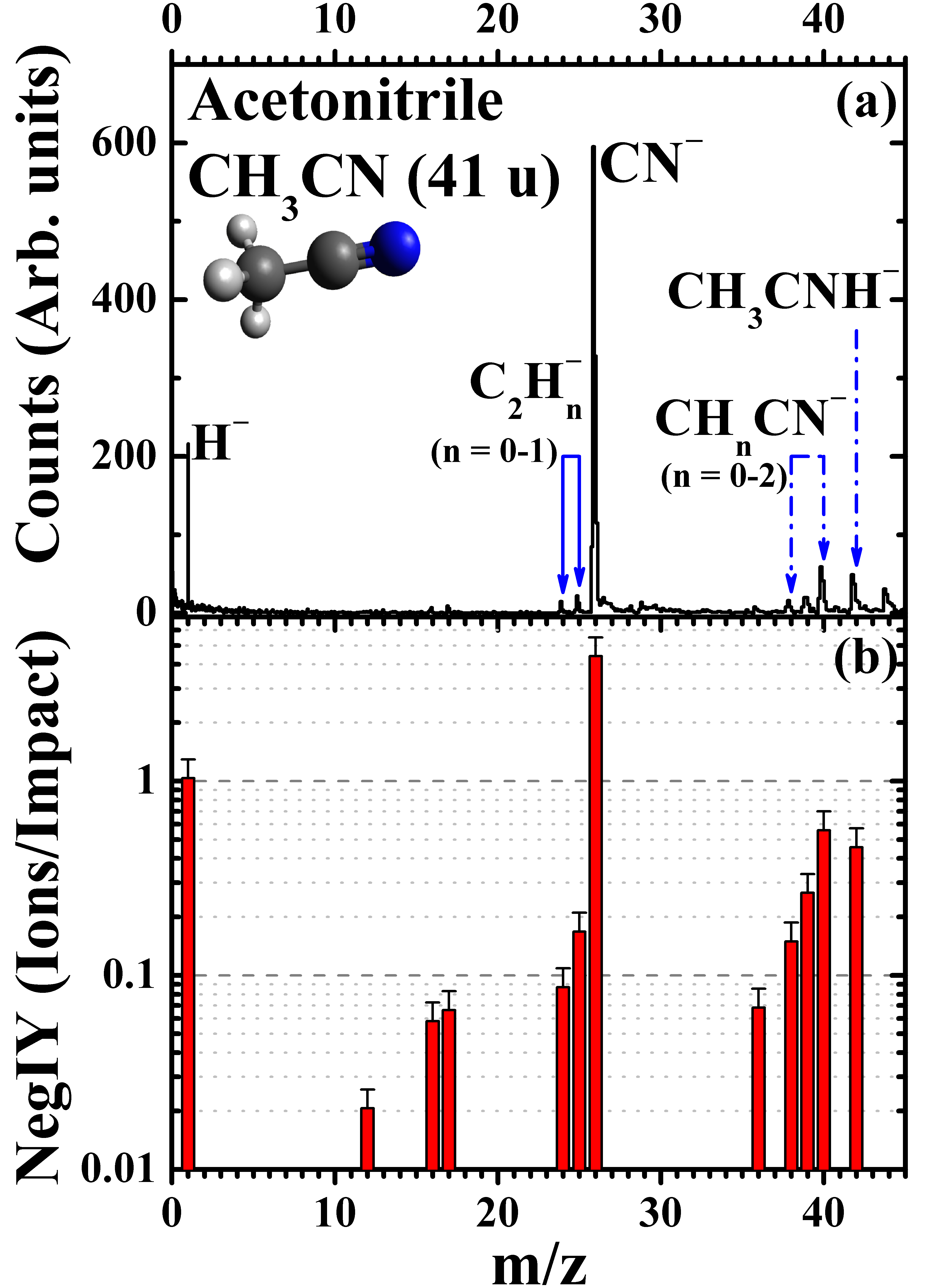}
    \caption{Top panel (a): Mass spectrum of negative secondary ions desorbed from the acetonitrile (CH$_3$CN) ice at 60 K. Bottom panel (b): Integrated NegIY (ions/impact) values for each desorbing fragment.}
    \label{Fig.6}
\end{figure}

\begin{figure}
	\includegraphics[width=\columnwidth]{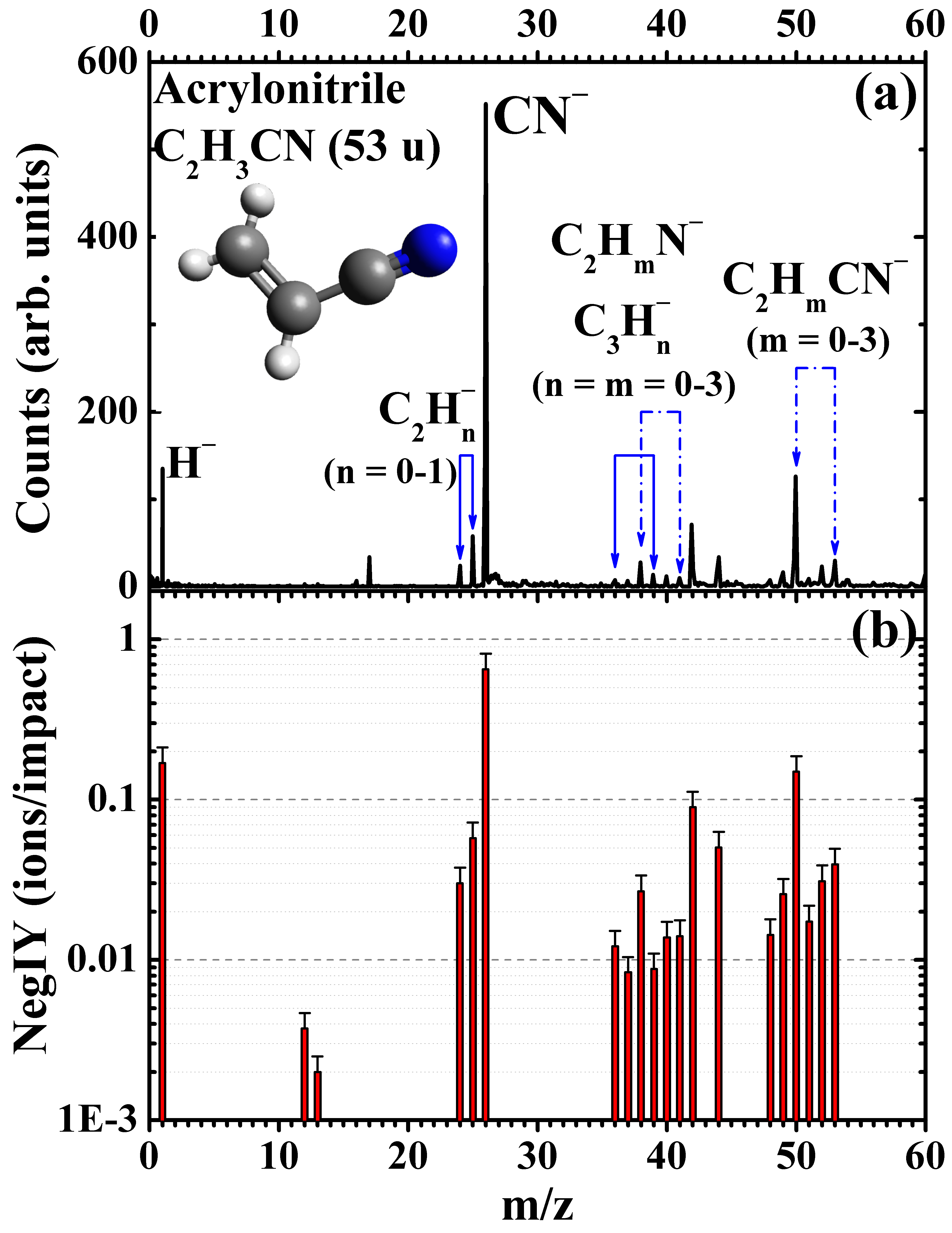}
    \caption{Top panel (a): Mass spectrum of negative secondary ions desorbed from the acrylonitrile ($C_2H_3CN$) ice at 60 K. Bottom panel (b): Integrated NegIY (ions/impact) values for each desorbing fragment.}
    \label{Fig.7}
\end{figure}

\begin{figure}
	\includegraphics[width=\columnwidth]{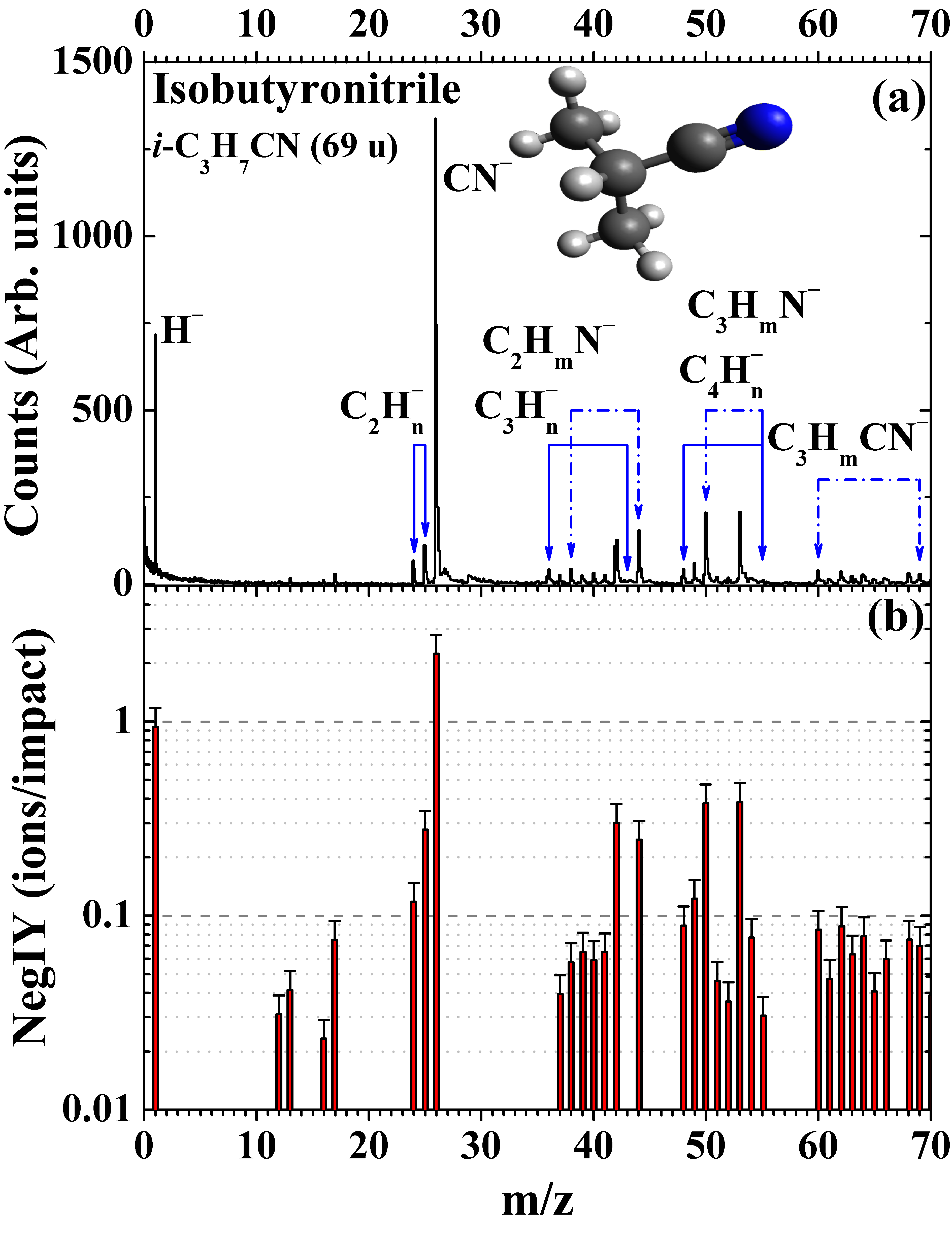}
    \caption{Top panel (a): Mass spectrum of negative secondary ions desorbed from the isobutyronitrile (i-C$_3$H$_7$CN) ice at 60 K. Bottom panel (b): Integrated NegIY (ions/impact) values for each desorbing fragment.}
    \label{Fig.8}
\end{figure}

The PDMS mass spectra of anion desorption from the CH$_3$CN, C$_2$H$_3$CN and i-C$_3$H$_7$CN ices are shown in the upper part (a) of Fig.~\ref{Fig.6}, Fig.~\ref{Fig.7} and Fig.~\ref{Fig.8}, respectively, while the negative ion yield (NegIY) for each desorbed fragment is graphically shown in the lower part (b). By comparing to the desorption yield of positive ions, one may note a considerable reduction in the yield of secondary negative ions. On the other hand, the NegIY values for individual ions are greater than the positive (PosIY) ones. Generally, desorption cross-sections for negative charged fragments are lower than those of cations, since there is secondary electron emission and because the positive secondary ions are strongly repelled by the positive infra-track \citep{Iza2007}.\\
The incidence of energetic heavy ions leads mainly to the ionisation of the target molecules along the projectile trajectory (infra-track), producing a cascade of secondary electrons ($\delta$-electrons) \citep{Martinez2012}. These $\delta$-electrons are produced with a range of electron energies and they may escape from the solid or diffuse almost radially from the infra-track\citep{Farenzena2006}. If the energy of the scattered $\delta$ electrons is low enough, they may attach to the molecules surrounding the ion track. The enhanced NegIY values in respect to the PosIY point to the formation of very stable anions, especially if one of the dissociating fragments has a positive electron affinity, resulting in a thermodynamically stable anion.\\

The high intensity m/z = 26 peak, attributed to CN$^-$ desorption, appears as a remarkable feature in the negative PDMS mass spectra from Fig.~\ref{Fig.6} to Fig.~\ref{Fig.8}. The desorption yield of this channel may be related to the energy involved in the formation of CN$^-$ from nitriles. Due to the similar R-CN bond dissociation energy for the aliphatic nitriles \citep{Blanksby2003}, and the relative high electron affinity of the CN radical (3.8620 eV) \citep{Bradforth1993}, desorption of CN$^-$ appears as the most energetically favoured dissociation channel. It is noteworthy that CN$^-$ desorption is also a prominent feature in the PDMS mass spectrum of negative ions desorbed from mixed CO-NH$_3$ ices at 25 K, where it has been argued that the CN$^-$ anion might be related to the production of the cyanate anion (OCN$^-$) \citep{Martinez2014}.\\
The relative NegIY (RNegIY) for the fragments at m/z = 24, 25 and 26 is presented in Fig.~\ref{Fig.8}. The RNegIY is shown as the ratio between the individual NegIY values and the NegIY sum over the 24 $\leq$ m/z $\leq$ 26 range. The RNegIY for CN$^-$ desorption has almost the same contribution for CH$_3$CN, C$_2$H$_3$CN and i-C$_3$H$_7$CN in the 24 $\leq$ m/z $\leq$ 26 range. Conversely, the RNegIYs for C$_{2}^{-}$ and C$_2$H$^-$ desorption seem to be sensitive to the extent of the alkyl chain.\\

\begin{figure}
	\includegraphics[width=\columnwidth]{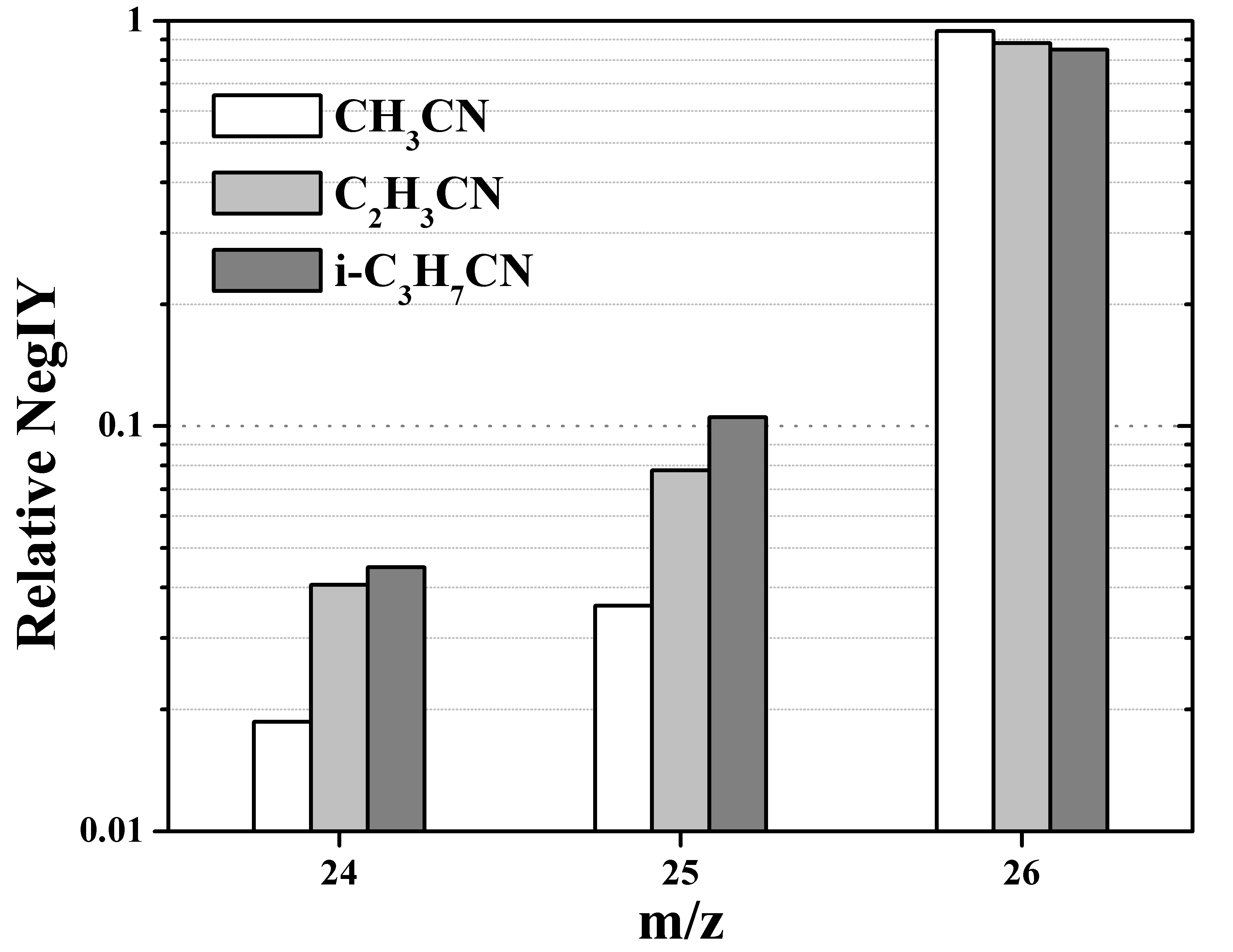}
    \caption{Relative NegIY for the anions in the 24 $\leq$ m/z $\leq$ 26 range for CH$_3$CN, C$_2$H$_3$CN and i-C$_3$H$_7$CN. Each individual NegIY was normalised by the summation over the total yield in the 24 $\leq$ m/z $\leq$ 26 range.}
    \label{Fig.9}
\end{figure}

\cite{Stockdale1974} observed two resonances for CN$^{-}$ and a high peak of CH$_2$CN$^{-}$ from CH$_3$CN in the gas-phase by low energy ($\leq$ 10 eV) electron impact, besides the evidence of CH$_3$CN$^-$ production. Two distinct energy regions (around 3 eV and between 6 and 12 eV) were also identified by \cite{Heni1986} for electron attachment and molecular dissociation of gas-phase CH$_3$CN, C$_2$H$_5$CN, i-C$_3$H$_7$CN and C$_2$H$_3$CN. Comparable results were also reported by \citep{Sailer2003}, in which the dominant anion observed from the gas-phase dissociation electron attachment (DEA) of CH$_3$CN is CH$_2$CN$^-$. Thus, though the gas-phase dissociation of nitriles exhibits a similar pattern to the present negative PDMS mass spectra, their relative intensities are clearly different.\\
\cite{Pirim2015} reported the cross-sections for electron stimulated desorption (ESD) of H$^-$ by the incidence of low-energy electrons onto amorphous and crystalline CH$_3$CN, condensed at 90 K and 130 K, respectively. Besides H$^-$ desorption, \cite{Bass2012} also reported low-energy ESD of CH$_2^-$, CH$_3^-$ and CN$^-$ from multilayer CH$_3$CN deposited onto Pt at 35 K. In contrast to the gas-phase DEA experiments, the authors reported only weak signals of CHCN$^-$ and CH$_2$CN$^-$ desorption. It is noteworthy that the $^{252}$Cf-PDMS spectrum of CH$_3$CN shown in Fig.~\ref{Fig.6}a presents only a minor desorption yield for CHCN$^-$ and CH$_2$CN$^-$, and the absence of CH$_3$CN$^-$ desorption. This fact is probably related to the production of short lifetime anions combined to a lower desorption probability for heavier fragments.\\
Replacement of the aliphatic chain from methyl to vinyl and isopropyl leads to desorption enhancement of C$^-$, besides the appearance of the C$_x$H$_n^-$ and C$_{x-1}$H$_m$N$^-$ ion series. Some few examples include the peaks observed at m/z = 38, 50 and 62, indicating the ion desorption of C$_2$N$^-$, C$_3$N$^-$ and C$_4$N$^-$, respectively. The relative intensities within each series might be related to the relative high electron affinity of the respective neutral radicals, on the order of ~ 2 up to ~ 4 eV \citep{Garand2009}.\\
On the other hand, the C$_2$H$_3$CN$^-$ parent ion is observed at m/z = 53 for acrylonitrile (Fig.~\ref{Fig.7}a). Due to its negative electron affinity, the mechanism leading to C$_2$H$_3$CN$^-$ desorption is challenging. It is known that isolated acrylonitrile does not lead to stable anions \citep{Maloteau1992}. However, because of its high dipole moment ($\mu$ = 3.87 D), acrylonitrile can form a dipole-bound negative ion, whose measured binding energy is 6.505 meV \citep{Suess2003}. Since the attachment of electrons with energy $\geq$ 2 eV is purely dissociative \citep{Heni1986}, C$_2$H$_3$CN$^-$ desorption might involve $\delta$-electrons of very low energy.

\subsection{Desorption of positive and negative ionic clusters}

\begin{figure}
	\includegraphics[width=\columnwidth]{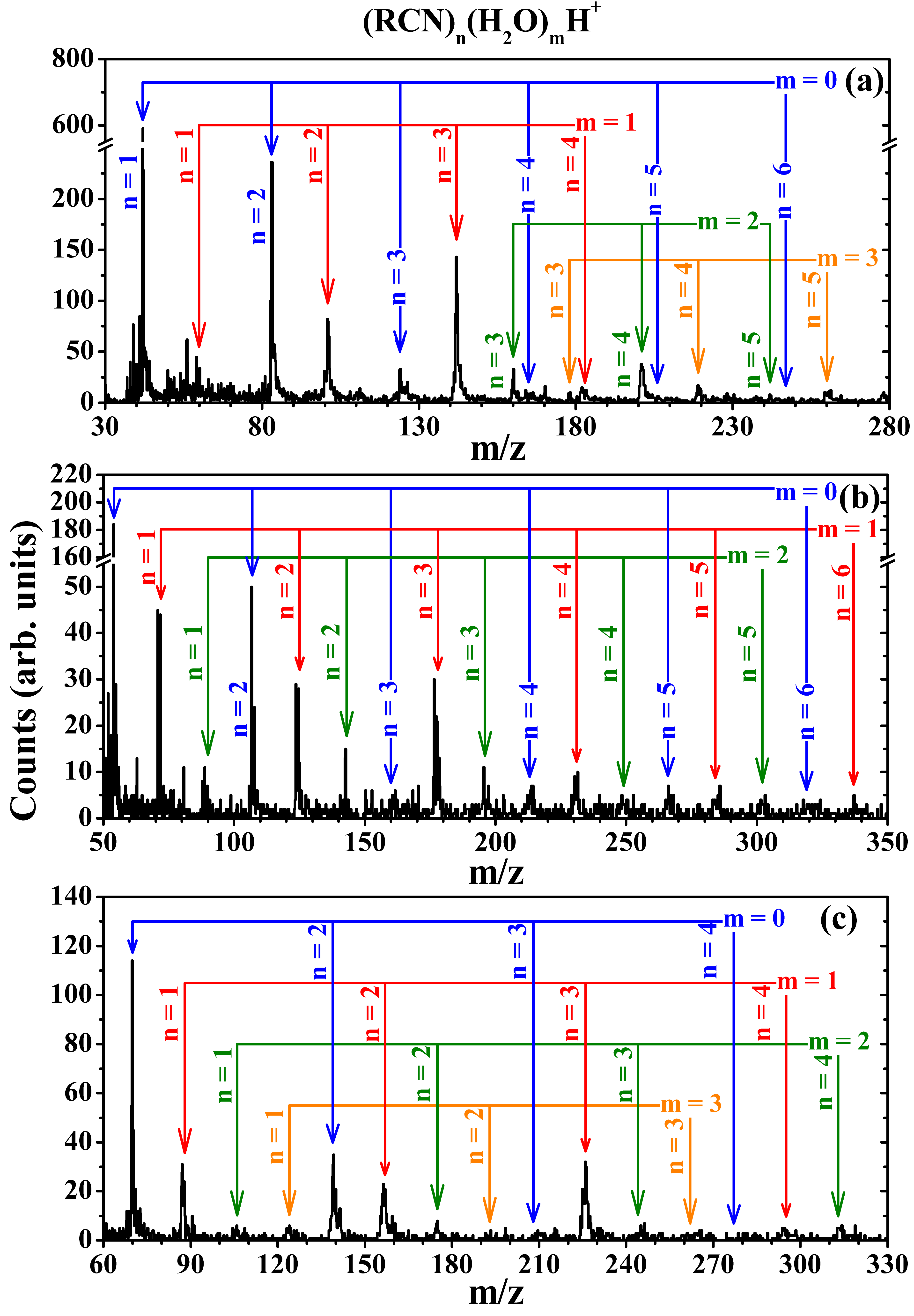}
    \caption{$^{252}$Cf-PDMS mass spectra of (RCN)$_n$(H$_2$O)$_m$H$^{+}$ cluster ions desorbed from CH$_3$CN (a), C$_2$H$_3$CN (b), and i-C$_3$H$_7$CN (c) ices at 60 K. Each series is designated by \textit{m} molecules of water (within the cluster). Color online.}
    \label{Fig.10}
\end{figure}

The processing of CH$_3$CN ice by 200 keV \citep{Abdulgalil2013} and 0.8 MeV \citep{Hudson2008} proton bombardment is known to produce new molecular species, including HCCCN and the isomerisation into isonitrile (CH$_3$NC) and ketenimine (H$_2$CCNH). Though such molecules are produced within the ice bulk, evidence for their formation may be achieved by ion desorption. Whether CH$_3$NC$^{+}$ or H$_2$CCNH$^{+}$ are produced along with CH$_3$CN$^{+}$ (m/z = 41), they cannot be resolved by the present mass spectrometer. However, production of HCCCN from pure CH$_3$CN requires further association between dissociating molecules. Accordingly, \cite{Hudson2008} identified features corresponding to succinonitrile (NCCH$_2$CH$_2$CN) in the infrared spectra of CH$_3$CN ice processed by 0.8 MeV protons. The authors proposed the formation of the later by dimerisation of CH$_2$CN radicals within the ice.\\
Previous PDMS experiments with H$_2$O, CO, CO$_2$, NH$_3$, HCOOH and CH$_3$OH \citep{Collado2004,Martinez2006,Andrade2007,Andrade2009} demonstrated a high tendency of clustering by $^{252}$Cf-FF bombardment. In the current experiments with nitriles, several ions were identified for each nitrile beyond the m/z ratio of the protonated parent molecule (RCNH$^{+}$), as presented in Fig~\ref{Fig.10}. Co-adsorption of water vapor from the residual gas is seen to play a role in cluster desorption, since the stoichiometry of positive ion clusters can be generalised by the formula (RCN)$_n$(H$_2$O)$_m$H$^{+}$. Though the presence of water is an undesired issue, its presence is of valuable information for astrophysical ices, since water is the most abundant constituent of ice mantles.\\

\begin{figure}
	\includegraphics[width=\columnwidth]{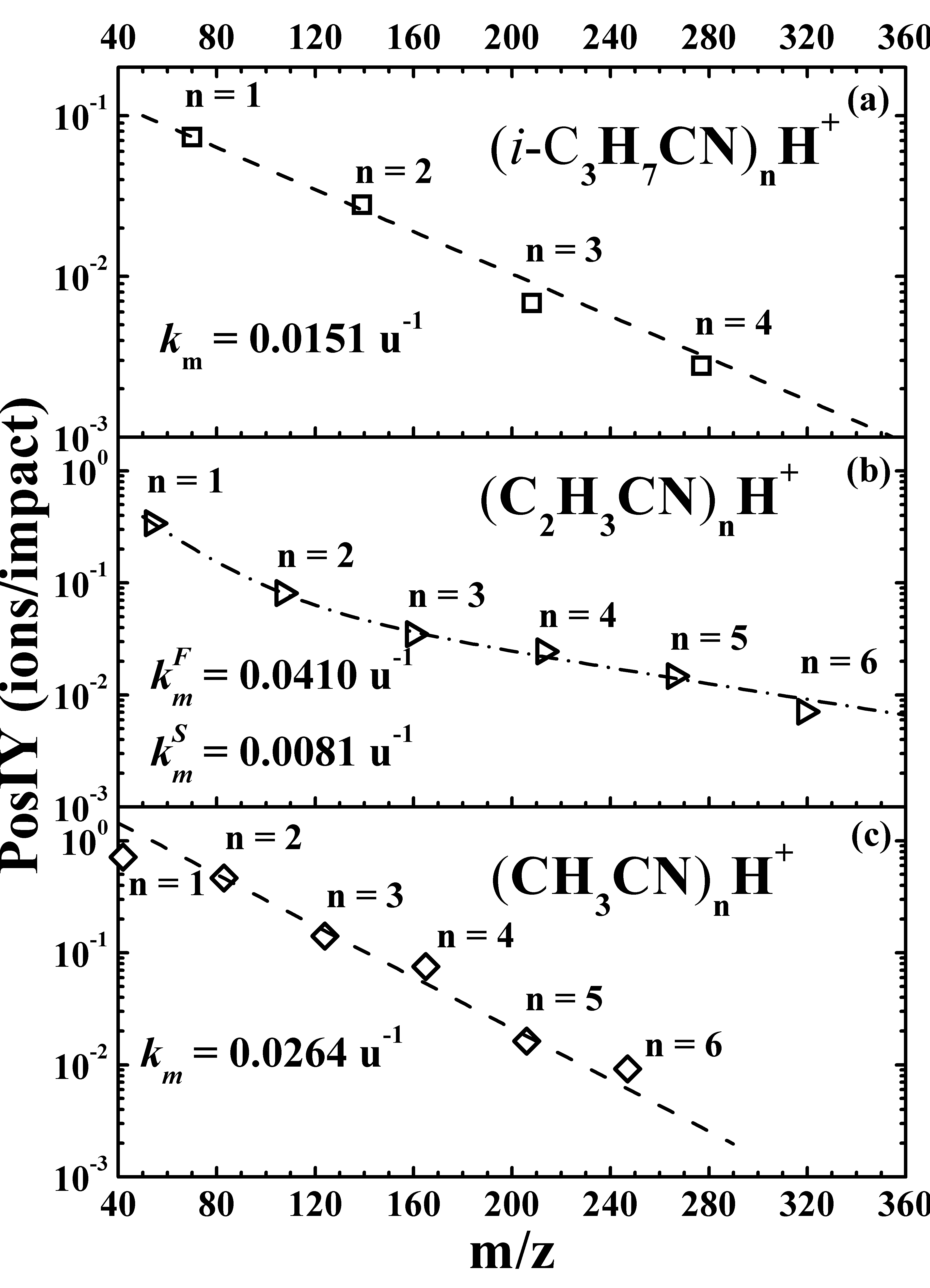}
    \caption{Positive ion yield (PosIY) as a function of the m/z ratio for the protonated nitrile clusters (RCN)$_n$H$^{+}$. The exponential slope constant, $k_m$, $k_{m}^{F}$ and $k_{m}^{S}$ for (a) i-C$_3$H$_7$CN, (b) C$_2$H$_3$CN and (c) CH$_3$CN are determined depending whether the data are fitted by a single or double exponential function.}
    \label{Fig.11}
\end{figure}

The presence of water molecules within the nitrile clusters indicates that water intermolecular interactions might be replaced by nitrile-water interactions, resulting in a solvating structure. This seems to enhance the production of protonated clusters of higher nuclearity (higher \textit{n} values), as seen by inspection of Fig.~\ref{Fig.10}. Each distribution is denoted by the number of water molecules (\textit{m}) within the cluster, and its maximum is systematically dependent on the number (\textit{n}) of nitriles interacting with them. For example, while a protonated cluster composed of two CH$_3$CN molecules possess a higher yield when no water is present (m/z = 83), three CH$_3$CN molecules desorb more efficiently whether bound to one water molecule (m/z = 142), and the same is true for four CH$_3$CN molecules bounded to two water molecules (m/z = 201). As can be noted in Fig.~\ref{Fig.10}a, these ions can be generalised by the formula (CH$_3$CN)$_{m+2}$(H$_2$O)$_{m}$H$^{+}$. A similar trend was reported by \cite{Wakisaka1998} on the analysis of hydrated clusters isolated from liquid droplets of acetonitrile aqueous solutions. They observed a reduced number of (CH$_3$CN)$_n$H$^{+}$ clusters from acetonitrile-water mixtures in comparison to the pure liquid acetonitrile. \cite{Wakisaka1998} concluded that the dipole-dipole interactions within CH$_3$CN clusters are reduced in the presence of water, disintegrating the CH$_3$CN clusters. Both in solid and liquid phases, water is seen to control the extent to which the hydrated clusters grow.\\
Despite the high proton affinities of nitriles, theoretical calculations made by \cite{Mejias2001} for the solvation of H$_3$O$^{+}$ in acetonitrile in the form of (CH$_3$CN)$_n$H$_3$O$^{+}$ (n = 1-3) clusters showed that a bare proton remains bounded to acetonitrile only when n = 1, leading to the CH$_3$CNH$^{+}$–H$_2$O complex. For n = 2 and 3, the optimised structures revealed that the proton remains as solvated H$_3$O$^{+}$, coordinating the CH$_3$CN molecules toward itself on a quasi-planar configuration. It is noteworthy that the protonated nitriles CH$_3$CNH$^{+}$, C$_2$H$_3$CNH$^{+}$ and i-C$_3$H$_7$CNH$^{+}$ are produced very efficiently from their respective ices, but the desorption yield of their (RCN)$_n$H$^{+}$ series steeply falls as the nuclearity (\textit{n}) increases. Fig.~\ref{Fig.11} presents the yield curves for the (RCN)$_n$H$^{+}$ ionic cluster series as a function of the m/z ratio for i-C$_3$H$_7$CN (a), C$_2$H$_3$CN (b) and CH$_3$CN (c) ices, respectively. The yield curves in Fig.~\ref{Fig.11}a and Fig.~\ref{Fig.11}c were fitted by a single decay exponential function, $Y=Y_{0}e^{(-k_{m}m/z)}$, whereas the yield curve in Fig.~\ref{Fig.11}b is better fitted by a combination of two exponential decay curves, $Y=Y_{0}^{F}e^{(-k_{m}^{F}m/z)} + Y_{0}^{S}e^{(-k_{m}^{S}m/z)}$.\\

The exponential decay of the yield of the desorbed ionic clusters has already been reported in the literature for other ices bombarded by $^{252}$Cf-FF, including H$_2$O, CO, CO$_2$, NH$_3$ and CH$_3$COH \citep{Collado2004,Farenzena2006,Ponciano2006,Martinez2006,Andrade2009}. It is generally accepted that the double exponential behaviour may be described by a cluster emission model based on two regimes, the so-called fast (F-regime) and slow (S-regime) regimes \citep{Martinez2006}.\\
Since H$^{+}$ desorbs mainly from the impact site \citep{Iza2006} and is by far one of the most probable fragments to desorb, ion-molecule interactions in the expanding plasma may lead to (RCN)$_n$H$^{+}$ clustering. On the other hand, the S-regime typically shows a dependence on the ice temperature and bulk structure, occurring by fragmentation of the excited solid on a colder region around the infra-track \citep{Martinez2006}. Thus, the pre-formed clusters on the surface might also fragment and desorb. Nonetheless, one has also to take into account that neutralisation on the surface turns out to be more efficient for larger m/z values, thus quenching desorption of clusters of high pre-formed nuclearity.\\
Considering the relatively small amount of material ejected into vacuum, the production of large (RCN)$_n$H$^{+}$ clusters in the current PDMS measurements is seen to agree with previous experiments with supersonic beams. \cite{Knoezinger1993} concluded that the occurrence of the (CH$_3$CN)$_n$H$^{+}$ ion series by electron impact in supersonic beams is dependent on the degree of aggregation \textit{n} of the neutral (CH$_3$CN)$_n$ cluster in the expanding gas. At very low concentrations (diluted CH$_3$CN in the carrier gas), the monomer molecules prevail in the beam, thus only the (CH$_3$CN)$_n$H$^{+}$ clusters with 1 $\leq$ \textit{n} $\leq$ 3 were identified. In the case of consecutive reactions of CH$_3$CNH$^{+}$ ions with CH$_3$CN in high-pressure gas flow experiments, the rate coefficients to produce (CH$_3$CN)$_2$H$^{+}$ and (CH$_3$CN)$_3$H$^{+}$ are known to systematically decrease, thus leading to a very slow reaction to produce the cluster with n = 3 \citep{Plasil1999}. Thus, association reactions between desorbing protons and nitriles in the expanding plasma after $^{252}$Cf-FF ion bombardment are characterised by the relatively high $k_{m}^{F}$  decay constant.\\
The ion yield dependence can also be described in respect to the cluster nuclearity by taking $k_{n}=m_{RCN}k_{m}$, where $m_{RCN}$ is the molecule mass in atomic mass units. As pointed out by \cite{Farenzena2006}, the $k_{n}^{F}$, $k_{n}^{S}$ and $k_{n}$ decay constants expressed in terms of the cluster nuclearity are more appropriate for describing atomic rearrangements. The parameters obtained by fitting the ion cluster yield curves in Fig.~\ref{Fig.11} are presented and compared to the decay constant derived in the $^{252}$Cf-PDMS mass spectrum for selected H$_2$O, CO, CO$_2$, NH$_3$ and CH$_3$OH clusters in Table~\ref{table2}.\\
    
\begin{table}
	\centering
	\caption{Decay constants $k_m$, $k_m^F$ and $k_m^S$ for the protonated nitrile clusters, (RCN)$_n$H$^+$, at 60 K derived from Fig.~\ref{Fig.11} and compared to typical values available in the literature.}
	\label{table2}
	\begin{tabular}{lccr} 
		\hline
	Cluster & $k_{m}^{F}$ (u$^{-1}$) & $k_{m}^{S}$ (u$^{-1}$) & $k_m$ (u$^{-1}$)\\
	\hline
	(CH$_3$CN)$_n$H$^{+}$ & - & - & 0.0264\\
	(C$_2$H$_3$CN)$_n$H$^{+}$ & 0.0410 & 0.0081 & -\\
	(i-C$_3$H$_7$CN)$_n$H$^{+}$ & - & - & 0.0151\\
	(H$_2$O)$_n$H$^{+}$ & 0.045$^{\textit{a}}$ & 0.008$^{\textit{a}}$ & -\\
	(CO)$_{n}^{+}$ & - & - & 0.007$^{\textit{b}}$\\
	(CO)$_n$CO$^{+}$ & - & - & 0.006$^{\textit{c}}$\\
	(CO)$_n$CO$_{2}^{+}$ & - & - & 0.007$^{\textit{c}}$\\
	(NH$_3$)$_n$NH$_{4}^{+}$ & 0.036$^{\textit{d}}$ & 0.007$^{\textit{d}}$ & -\\
	(CH$_3$OH)$_n$CH$_3$OH$_{2}^{+}$ & - & - & 0.013$^{\textit{e}}$\\
		\hline
	\end{tabular}
	
	\medskip
    $^{\textit{a}}$ \cite{Collado2004}; $^{\textit{b}}$ \cite{Farenzena2006}; $^{\textit{c}}$ \cite{Ponciano2006}; $^{\textit{d}}$ \cite{Martinez2006}; $^{\textit{e}}$ \cite{Andrade2009}.
\end{table}

\begin{figure}
	\includegraphics[width=\columnwidth]{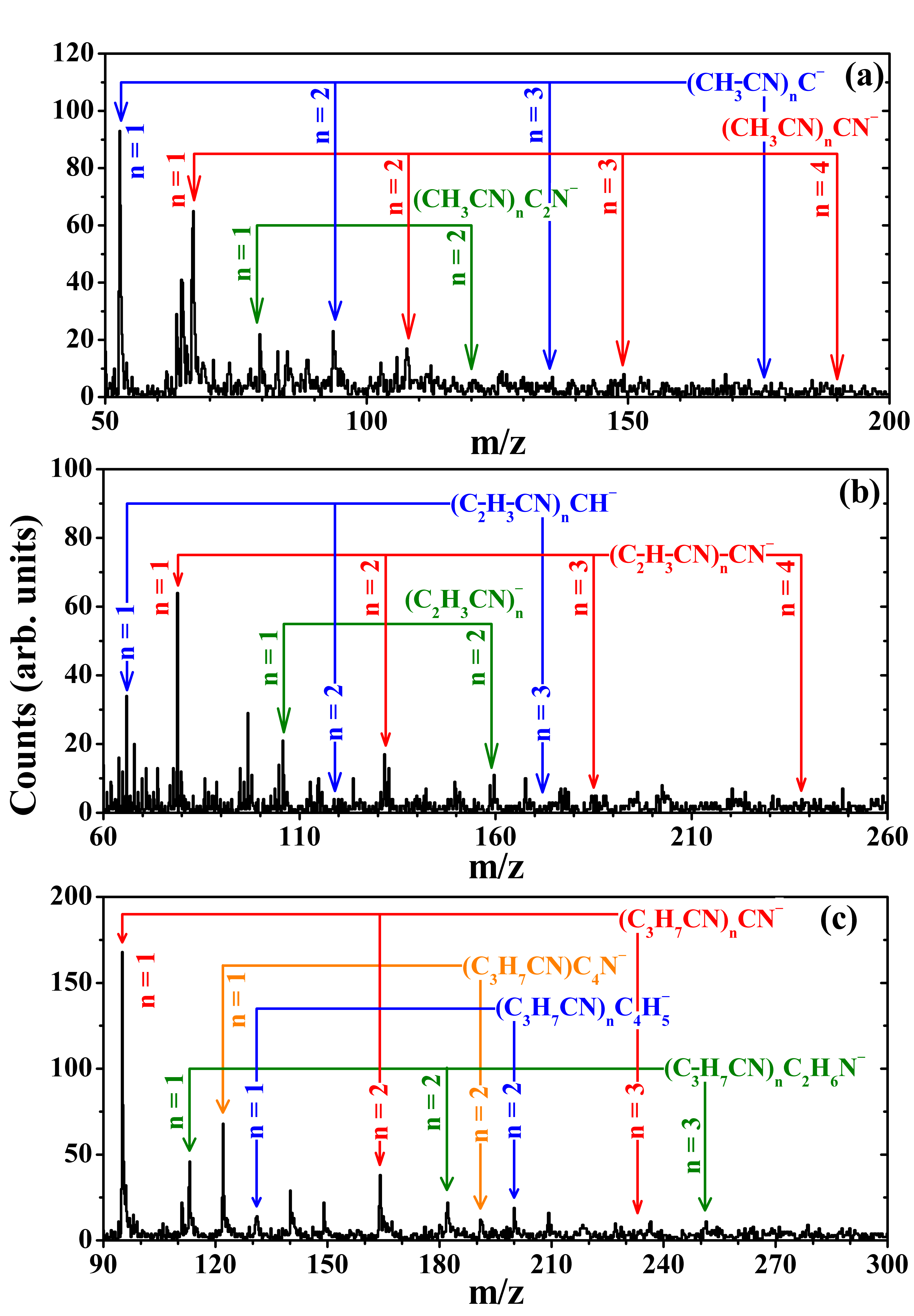}
    \caption{$^{252}$Cf-PDMS mass spectra of (RCN)$_n$CN$^{-}$ cluster anions desorbed from (a) CH$_3$CN, (b) C$_2$H$_3$CN and (c) i-C$_3$H$_7$CN ices at 60 K. Color online.}
    \label{Fig.12}
\end{figure}

It is noteworthy that the $k_{m}^{F}$ and $k_{m}^{S}$ constants derived for (C$_2$H$_3$CN)$_n$H$^+$ are almost on the same magnitude as those derived for (H$_2$O)$_n$H$^+$ and (NH$_3$)$_n$NH$_{4}^+$ ionic clusters. The same is valid for the single decay constant $k_m$ derived for (CH$_3$CN)$_n$H$^+$ and (i-C$_3$H$_7$CN)$_n$H$^+$ in comparison to (CH$_3$OH)$_n$CH$_3$OH$_{2}^+$. The last is not true for ion clusters bearing CO and CO$_2$ molecules. Since the $k_m^F$, $k_m^S$ and $k_m$ decay constants are defined in respect to the cluster mass, they should be sensitive to desorption probabilities and, thus, such deviation may point out to different neutralisation times. This may also be rationalised by taking into account that CO and CO$_2$ are more volatile than nitriles, thus ejecting more neutral molecules from the surface during the bulk sublimation on the infra-track.\\
The desorption of negative ionic clusters appears more complex than the respective positive ionic clusters. The negatively charged clusters observed in the $^{252}$Cf-PDMS mass spectrum of CH$_3$CN, C$_2$H$_3$CN and i-C$_3$H$_7$CN are shown in Fig.~\ref{Fig.12}a–c, respectively. Whereas the positive ionic cluster spectrum is dominated by hydrated and protonated clusters, many combinations between dissociating fragments and neutral molecules were observed to produce anion clusters.\\

\begin{figure}
	\includegraphics[width=\columnwidth]{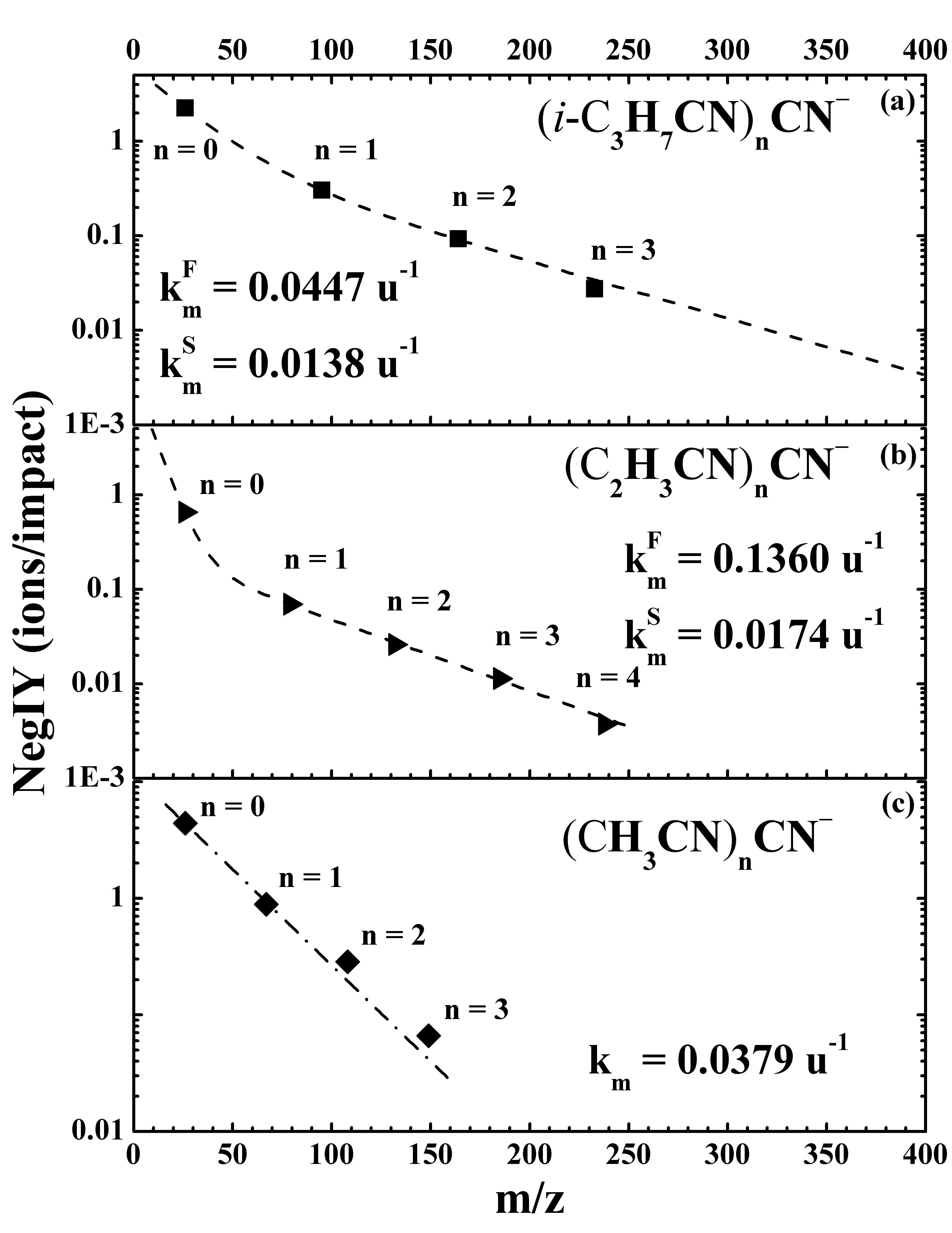}
    \caption{Negative ion yield (NegIY) as a function of the m/z ratio for nitrile clusters (RCN)$_n$CN$^-$. The exponential slope constants, $k_m$, $k_{m}^{F}$ and $k_{m}^{S}$ for (a) i-C$_3$H$_7$CN, (b) C$_2$H$_3$CN and (c) CH$_3$CN were determined depending on whether the data was fitted by a single or double exponential function.}
    \label{Fig.13}
\end{figure}

For example, a strong peak at m/z = 53 and the presence of m/z = 94 in Fig.~\ref{Fig.12}a indicate the desorption of the (CH$_3$CN)$_n$C$^-$ (n = 1-2) clusters, though no appreciable peak at m/z = 12 (C$^-$) was observed for CH$_3$CN (Fig.~\ref{Fig.2}a). Peaks at m/z = 60 and m/z = 119 for C$_2$H$_3$CN (Fig.~\ref{Fig.12}b) are indicative of (C$_2$H$_3$CN)$_n$CH$^-$ (n = 1-2) desorption. For the isobutyronitrile ice, the cluster series (i-C$_3$H$_7$CN)$_n$C$_2$H$_3$CN$^-$ (n = 1-2) is also suggested due to the presence of peaks at m/z = 122 and m/z = 191. Radicals of high-electron affinity, such as C$_2$N$^-$ and C$_4$N$^-$, also form the clusters (CH$_3$CN)$_n$C$_2$N$^-$ and (i-C$_3$H$_7$CN)$_n$C$_4$N$^-$. Then, one of the main cluster series desorbed is attributed to (RCN)$_n$CN$^-$.\\
Analogous to the positive ionic clusters (RCN)$_n$H$^+$, Fig.~\ref{Fig.13}a–c presents the yield curves for the anion cluster series (RCN)$_n$CN$^-$ as a function of the m/z ratio for (a) i-C$_3$H$_7$CN, (b) C$_2$H$_3$CN and (c) CH$_3$CN ices. The yield curves presented were fitted by the sum of two exponential decay functions, $Y=Y_{0}^{F}e^{(-k_{m}^{F}m/z)} + Y_{0}^{S}e^{(-k_{m}^{S}m/z)}$, except for CH$_3$CN, in which a single decay exponential function, $Y=Y_{0}e^{(-k_{m}m/z)}$, fitted the data. In this case, $Y_0$ denotes the ion yield for n = 0, i.e., the NegIY for CN$^-$. The two exponentials in the former account for the fast (F) and slow (S) regime for cluster formation. The parameters obtained by fitting the ion cluster yield curves in Fig.~\ref{Fig.13} are presented in Table 3 and compared to the decay constant derived in the $^{252}$Cf-PDMS mass spectrum for selected H$_2$O, CO$_2$, and NH$_3$ negative ionic clusters from the literature.\\

\begin{table}
	\centering
	\caption{Decay constants $k_m$, $k_{m}^{F}$ and $k_{m}^{S}$ for the negative ionic clusters, (RCN)$_n$CN$^{-}$, at 60 K derived from Fig.~\ref{Fig.13} and compared to typical values available in the literature.}
	\label{table3}
	\begin{tabular}{lccr} 
		\hline
	Cluster & $k_{m}^{F}$ (u$^{-1}$) & $k_{m}^{S}$ (u$^{-1}$) & $k_m$ (u$^{-1}$)\\
	\hline
	(CH$_3$CN)$_n$CN$^{-}$ & - & - & 0.0379\\
	(C$_2$H$_3$CN)$_n$CN$^{-}$ & 0.1360 & 0.0174 & -\\
	(i-C$_3$H$_7$CN)$_n$CN$^{-}$ & 0.0447 & 0.0138 & -\\
	(H$_2$O)$_n$OH$^{-}$ & - & - & 0.007$^{\textit{a}}$\\
	(CO$_2$)$_{n}^{-}$ & - & - & 0.02$^{\textit{b}}$\\
	(CO$_2$)$_n$CO$_{3}^{-}$ & - & - & $^{\textit{c}}$\\
	(NH$_3$)$_n$NH$_{2}^{-}$ & 0.039$^{\textit{d}}$ & 0.007$^{\textit{d}}$ & -\\
		\hline
	\end{tabular}
	
	\medskip
    $^{\textit{a}}$ \cite{Collado2004}; $^{\textit{b}}$ \cite{Farenzena2006}; $^{\textit{c}}$ \cite{Ponciano2006}; $^{\textit{d}}$ \cite{Martinez2006}.
\end{table}

Even though the $k_m$ decay constants for (RCN)$_n$CN$^-$ clusters are comparable, the $k_m$  values for anions are almost three times those derived for (RCN)$_n$H$^+$ clusters. This increase may be correlated to a more efficient neutralisation mechanism for large anions prior to desorption. Also, as evaluated by their respective decay constants $k_m$, (RCN)$_n$CN$^-$ clusters of large nuclearity are not expected to desorb efficiently, as compared to water and other volatile species that own considerably lower $k_m$ values.\\

\section{Conclusions}

Heavy-ion bombardment onto nitrile-bearing ices showed a pronounced desorption yield for protonated nitriles, along with the desorption of protonated clusters (RCN)$_n$H$^+$ of nuclearity extending up to six (n $\leq$ 6). This indicates that molecular rearrangement plays a role in desorption induced by heavy ion bombardment. A similar mechanism for ion desorption is expected by the relaxation dynamics of the solid after the passage of a MeV-cosmic-ray onto interstellar/interplanetary ices.\\
The presence of cations, in particular of protonated molecules, in the gas phase is essential for producing neutrals in the ISM, since the latter may result from the dissociative recombination of positive ions. In this sense, desorption of CH$_3$CNH$^+$, C$_2$H$_3$CNH$^+$, i-C$_3$H$_7$CNH$^+$ and the respective (RCN)$_n$H$^+$ clusters from ices by heavy ion impact is expected to play a role in determining the abundance of the respective neutrals. Cosmic-ray bombardment of ices bearing nitriles would thus contribute to the delivery of nitriles and related molecules (such as amines) to the gas phase in cold regions, where thermal desorption is negligible.

Anion desorption following heavy ion bombardment showed an enhanced efficiency in comparison to positive ion desorption. Heavy ion bombardment onto CH$_3$CN, C$_2$H$_3$CN and i-C$_3$H$_7$CN ices strongly induces the desorption of CN$^-$, as well as other anions attributed to the C$_n$H$_{x}^{-}$, CH$_m$CN$^-$ ion series and the (RCN)$_n$CN$^-$ cluster series. Few examples among these species include the relevant anions C$_2$N$^-$, C$_3$N$^-$ and C$_4$N$^-$, which are important for the growth of cyanopolyynes. A favourable argument relies on the relative high electron affinity of the respective neutral radicals.\\
The relatively high proton affinities of nitriles and the considerable exoenergic electron affinity of the cyano group constitute two arguments that favour the desorption of protonated nitriles and negative ions from condensed nitriles. The results obtained here may be of fundamental importance to the balance of gas-phase species observed towards the ISM, especially in dark molecular clouds where thermal desorption is negligible. Thus, more complex nitriles and related compounds shall probably be detected in the ISM with the advance of observational facilities.\\

\section*{Acknowledgements}

The authors would like to acknowledge Conselho Nacional de Desenvolvimento Cientifico e Tecnológico (CNPq), Coordenação de Aperfeiçoamento de Pessoal de Nível Superior (CAPES), and Fundação de Amparo à Pesquisa do Estado do Rio de Janeiro (FAPERJ) for financial support. The authors also acknowledge the Van de Graaff staff for technical support during the experiments.



\bibliographystyle{mnras}
\bibliography{Ref}

\begin{thebibliography}{}
\makeatletter
\relax
\def\mn@urlcharsother{\let\do\@makeother \do\$\do\&\do\#\do\^\do\_\do\%\do\~}
\def\mn@doi{\begingroup\mn@urlcharsother \@ifnextchar [ {\mn@doi@}
  {\mn@doi@[]}}
\def\mn@doi@[#1]#2{\def\@tempa{#1}\ifx\@tempa\@empty \href
  {http://dx.doi.org/#2} {doi:#2}\else \href {http://dx.doi.org/#2} {#1}\fi
  \endgroup}
\def\mn@eprint#1#2{\mn@eprint@#1:#2::\@nil}
\def\mn@eprint@arXiv#1{\href {http://arxiv.org/abs/#1} {{\tt arXiv:#1}}}
\def\mn@eprint@dblp#1{\href {http://dblp.uni-trier.de/rec/bibtex/#1.xml}
  {dblp:#1}}
\def\mn@eprint@#1:#2:#3:#4\@nil{\def\@tempa {#1}\def\@tempb {#2}\def\@tempc
  {#3}\ifx \@tempc \@empty \let \@tempc \@tempb \let \@tempb \@tempa \fi \ifx
  \@tempb \@empty \def\@tempb {arXiv}\fi \@ifundefined
  {mn@eprint@\@tempb}{\@tempb:\@tempc}{\expandafter \expandafter \csname
  mn@eprint@\@tempb\endcsname \expandafter{\@tempc}}}

\bibitem[\protect\citeauthoryear{Abdulgalil et~al.,}{Abdulgalil
  et~al.}{2013}]{Abdulgalil2013}
Abdulgalil A. G.~M.,  et~al., 2013, \mn@doi [Philosophical Transactions of the
  Royal Society of London, Series A: Physical Sciences and Engineering]
  {10.1098/rsta.2011.0586}, 371, 20110586

\bibitem[\protect\citeauthoryear{Andrade, Rocco, Boechat-Roberty, Iza,
  Martinez, Homem  \& da Silveira}{Andrade et~al.}{2007}]{Andrade2007}
Andrade D.,  Rocco M.,  Boechat-Roberty H.,  Iza P.,  Martinez R.,  Homem M.,
  da Silveira E.,  2007, \mn@doi [Journal of Electron Spectroscopy and Related
  Phenomena] {https://doi.org/10.1016/j.elspec.2006.12.003}, 155, 124

\bibitem[\protect\citeauthoryear{Andrade, Boechat-Roberty, Martinez, Homem, da
  Silveira  \& Rocco}{Andrade et~al.}{2009}]{Andrade2009}
Andrade D.,  Boechat-Roberty H.,  Martinez R.,  Homem M.,  da Silveira E.,
  Rocco M.,  2009, \mn@doi [Surface Science]
  {https://doi.org/10.1016/j.susc.2009.02.035}, 603, 1190

\bibitem[\protect\citeauthoryear{Araya, Hofner, Kurtz, Bronfman  \&
  DeDeo}{Araya et~al.}{2005}]{Araya2005}
Araya E.,  Hofner P.,  Kurtz S.,  Bronfman L.,   DeDeo S.,  2005, \mn@doi [The
  Astrophysical Journal Supplement Series] {10.1086/427187}, 157, 279

\bibitem[\protect\citeauthoryear{Bass, Bredehöft, Böhler, Sanche  \&
  Swiderek}{Bass et~al.}{2012}]{Bass2012}
Bass A.,  Bredehöft J.,  Böhler E.,  Sanche L.,   Swiderek P.,  2012, \mn@doi
  [European Physical Journal D: Atomic, Molecular and Optical Physics]
  {10.1140/epjd/e2012-20670-y}, 66, 53

\bibitem[\protect\citeauthoryear{Bell, Cernicharo, Viti, Marcelino, Palau,
  Esplugues  \& Tercero}{Bell et~al.}{2014}]{Bell2014}
Bell T.~A.,  Cernicharo J.,  Viti S.,  Marcelino N.,  Palau A.,  Esplugues
  G.~B.,   Tercero B.,  2014, \mn@doi [Astronomy and Astrophysics]
  {10.1051/0004-6361/201321872}, 564, A114

\bibitem[\protect\citeauthoryear{Belloche, Garrod, M\"uller, Menten, Comito  \&
  Schilke}{Belloche et~al.}{2009}]{Belloche2009}
Belloche A.,  Garrod R.~T.,  M\"uller H. S.~P.,  Menten K.~M.,  Comito C.,
  Schilke P.,  2009, \mn@doi [Astronomy and Astrophysics]
  {10.1051/0004-6361/200811550}, 499, 215

\bibitem[\protect\citeauthoryear{Belloche, Garrod, M{\"u}ller  \&
  Menten}{Belloche et~al.}{2014}]{Belloche2014}
Belloche A.,  Garrod R.~T.,  M{\"u}ller H. S.~P.,   Menten K.~M.,  2014,
  \mn@doi [Science] {10.1126/science.1256678}, 345, 1584

\bibitem[\protect\citeauthoryear{Betts, da Silveira  \& Schweikert}{Betts
  et~al.}{1995}]{Betts1995}
Betts R.,  da Silveira E.,   Schweikert E.,  1995, \mn@doi [International
  Journal of Mass Spectrometry and Ion Processes]
  {https://doi.org/10.1016/0168-1176(95)04180-S}, 145, 9

\bibitem[\protect\citeauthoryear{Bisschop, J\o{}rgensen, van Dishoeck  \& {de
  Wachter}}{Bisschop et~al.}{2007}]{Bisschop2007}
Bisschop S.~E.,  J\o{}rgensen J.~K.,  van Dishoeck E.~F.,   {de Wachter} E.
  B.~M.,  2007, \mn@doi [Astronomy and Astrophysics]
  {10.1051/0004-6361:20065963}, 465, 913

\bibitem[\protect\citeauthoryear{Blanksby \& Ellison}{Blanksby \&
  Ellison}{2003}]{Blanksby2003}
Blanksby S.~J.,  Ellison G.~B.,  2003, \mn@doi [Accounts of Chemical Research]
  {10.1021/ar020230d}, 36, 255

\bibitem[\protect\citeauthoryear{{B{\o}gelund}, {Barr}, {Taquet}, {Ligterink},
  {Persson}, {Hogerheijde}  \& {van Dishoeck}}{{B{\o}gelund}
  et~al.}{2019}]{Bogelund2019}
{B{\o}gelund} E.~G.,  {Barr} A.~G.,  {Taquet} V.,  {Ligterink} N. F.~W.,
  {Persson} M.~V.,  {Hogerheijde} M.~R.,   {van Dishoeck} E.~F.,  2019, arXiv
  e-prints, \href {https://ui.adsabs.harvard.edu/abs/2019arXiv190606156B} {p.
  arXiv:1906.06156}

\bibitem[\protect\citeauthoryear{Bradforth, Kim, Arnold  \& Neumark}{Bradforth
  et~al.}{1993}]{Bradforth1993}
Bradforth S.~E.,  Kim E.~H.,  Arnold D.~W.,   Neumark D.~M.,  1993, \mn@doi
  [The Journal of Chemical Physics] {10.1063/1.464244}, 98, 800

\bibitem[\protect\citeauthoryear{Bringa \& Johnson}{Bringa \&
  Johnson}{2003}]{Bringa2003}
Bringa E.~M.,  Johnson R.~E.,  2003, in Pirronello V.,  Krelowski J.,
  Manic{\`o} G.,  eds, Solid State Astrochemistry. Springer Netherlands,
  Dordrecht, pp 357--393

\bibitem[\protect\citeauthoryear{Burke \& Brown}{Burke \&
  Brown}{2010}]{Burke2010}
Burke D.~J.,  Brown W.~A.,  2010, \mn@doi [Physical Chemistry Chemical Physics]
  {10.1039/B917005G}, 12, 5947

\bibitem[\protect\citeauthoryear{Cazaux, Tielens, Ceccarelli, Castets, Wakelam,
  Caux, Parise  \& Teyssier}{Cazaux et~al.}{2003}]{Cazaux2003}
Cazaux S.,  Tielens A. G. G.~M.,  Ceccarelli C.,  Castets A.,  Wakelam V.,
  Caux E.,  Parise B.,   Teyssier D.,  2003, \mn@doi [The Astrophysical
  Journal] {10.1086/378038}, 593, L51

\bibitem[\protect\citeauthoryear{Collado, Farenzena, Ponciano, da Silveira  \&
  Wien}{Collado et~al.}{2004}]{Collado2004}
Collado V.,  Farenzena L.,  Ponciano C.,  da Silveira E.,   Wien K.,  2004,
  \mn@doi [Surface Science] {https://doi.org/10.1016/j.susc.2004.07.022}, 569,
  149

\bibitem[\protect\citeauthoryear{Collings \& McCoustra}{Collings \&
  McCoustra}{2012}]{Collings2012}
Collings M.,  McCoustra M.,  2012, \mn@doi [EAS Publications Series]
  {10.1051/eas/1258052}, 58, 315

\bibitem[\protect\citeauthoryear{Collings, Anderson, Chen, Dever, Viti,
  Williams  \& McCoustra}{Collings et~al.}{2004}]{Collings2004}
Collings M.~P.,  Anderson M.~A.,  Chen R.,  Dever J.~W.,  Viti S.,  Williams
  D.~A.,   McCoustra M. R.~S.,  2004, \mn@doi [Monthly Notices of the Royal
  Astronomical Society] {10.1111/j.1365-2966.2004.08272.x}, 354, 1133

\bibitem[\protect\citeauthoryear{{Costes}, {Naulin}  \& {Dorthe}}{{Costes}
  et~al.}{1990}]{Costes1990}
{Costes} M.,  {Naulin} C.,   {Dorthe} G.,  1990, Astronomy and Astrophysics,
  \href {https://ui.adsabs.harvard.edu/abs/1990A&A...232..270C} {232, 270}

\bibitem[\protect\citeauthoryear{Couturier-Tamburelli, Pi\'etri  \&
  Gudipati}{Couturier-Tamburelli et~al.}{2015}]{Couturier-Tamburelli2015}
Couturier-Tamburelli I.,  Pi\'etri N.,   Gudipati M.~S.,  2015, \mn@doi
  [Astronomy and Astrophysics] {10.1051/0004-6361/201425518}, 578, A111

\bibitem[\protect\citeauthoryear{Farenzena et~al.,}{Farenzena
  et~al.}{2005a}]{Farenzena2005b}
Farenzena L.~S.,  et~al., 2005a, \mn@doi [Earth, Moon, and Planets]
  {10.1007/s11038-006-9081-y}, 97, 311

\bibitem[\protect\citeauthoryear{Farenzena, Collado, Ponciano, da Silveira  \&
  Wien}{Farenzena et~al.}{2005b}]{Farenzena2005a}
Farenzena L.,  Collado V.,  Ponciano C.,  da Silveira E.,   Wien K.,  2005b,
  \mn@doi [International Journal of Mass Spectrometry]
  {https://doi.org/10.1016/j.ijms.2004.12.034}, 243, 85

\bibitem[\protect\citeauthoryear{Farenzena, Martinez, Iza, Ponciano, Homem, de
  Brito, da Silveira  \& Wien}{Farenzena et~al.}{2006}]{Farenzena2006}
Farenzena L.,  Martinez R.,  Iza P.,  Ponciano C.,  Homem M.,  de Brito A.~N.,
  da Silveira E.,   Wien K.,  2006, \mn@doi [International Journal of Mass
  Spectrometry] {https://doi.org/10.1016/j.ijms.2005.12.006}, 251, 1

\bibitem[\protect\citeauthoryear{Fontani, Pascucci, Caselli, Wyrowski, Cesaroni
   \& Walmsley}{Fontani et~al.}{2007}]{Fontani2007}
Fontani F.,  Pascucci I.,  Caselli P.,  Wyrowski F.,  Cesaroni R.,   Walmsley
  C.~M.,  2007, \mn@doi [Astronomy and Astrophysics]
  {10.1051/0004-6361:20077485}, 470, 639

\bibitem[\protect\citeauthoryear{Gaches, Offner  \& Bisbas}{Gaches
  et~al.}{2019}]{Gaches2019}
Gaches B. A.~L.,  Offner S. S.~R.,   Bisbas T.~G.,  2019, \mn@doi [The
  Astrophysical Journal] {10.3847/1538-4357/ab20c7}, 878, 105

\bibitem[\protect\citeauthoryear{Gans, Boyé-Péronne, Garcia, Röder,
  Schleier, Halvick  \& Loison}{Gans et~al.}{2017}]{Gans2017}
Gans B.,  Boyé-Péronne S.,  Garcia G.~A.,  Röder A.,  Schleier D.,  Halvick
  P.,   Loison J.-C.,  2017, \mn@doi [The Journal of Physical Chemistry
  Letters] {10.1021/acs.jpclett.7b01853}, 8, 4038

\bibitem[\protect\citeauthoryear{Garand, Yacovitch  \& Neumark}{Garand
  et~al.}{2009}]{Garand2009}
Garand E.,  Yacovitch T.~I.,   Neumark D.~M.,  2009, \mn@doi [The Journal of
  Chemical Physics] {10.1063/1.3076320}, 130, 064304

\bibitem[\protect\citeauthoryear{{Garrod}}{{Garrod}}{2013}]{Garrod2013}
{Garrod} R.~T.,  2013, \mn@doi [The Astrophysical Journal]
  {10.1088/0004-637x/765/1/60}, 765, 60

\bibitem[\protect\citeauthoryear{Garrod, Weaver  \& Herbst}{Garrod
  et~al.}{2008}]{Garrod2008}
Garrod R.~T.,  Weaver S. L.~W.,   Herbst E.,  2008, \mn@doi [The Astrophysical
  Journal] {10.1086/588035}, 682, 283

\bibitem[\protect\citeauthoryear{Garrod, Belloche, M\"uller  \& Menten}{Garrod
  et~al.}{2017}]{Garrod2017}
Garrod R.~T.,  Belloche A.,  M\"uller H. S.~P.,   Menten K.~M.,  2017, \mn@doi
  [Astronomy and Astrophysics] {10.1051/0004-6361/201630254}, 601, A48

\bibitem[\protect\citeauthoryear{Heni \& Illenberger}{Heni \&
  Illenberger}{1986}]{Heni1986}
Heni M.,  Illenberger E.,  1986, \mn@doi [International Journal of Mass
  Spectrometry and Ion Processes]
  {https://doi.org/10.1016/0168-1176(86)80014-3}, 73, 127

\bibitem[\protect\citeauthoryear{Herbst \& Cuppen}{Herbst \&
  Cuppen}{2006}]{Herbst2006}
Herbst E.,  Cuppen H.~M.,  2006, \mn@doi [Proceedings of the National Academy
  of Sciences] {10.1073/pnas.0601556103}, 103, 12257

\bibitem[\protect\citeauthoryear{{Hudson} \& {Moore}}{{Hudson} \&
  {Moore}}{2004}]{Hudson2004}
{Hudson} R.~L.,  {Moore} M.~H.,  2004, \mn@doi [Icarus]
  {https://doi.org/10.1016/j.icarus.2004.06.011}, 172, 466

\bibitem[\protect\citeauthoryear{Hudson, Moore, Dworkin, Martin  \&
  Pozun}{Hudson et~al.}{2008}]{Hudson2008}
Hudson R.~L.,  Moore M.~H.,  Dworkin J.~P.,  Martin M.~P.,   Pozun Z.~D.,
  2008, \mn@doi [Astrobiology] {10.1089/ast.2007.0131}, 8, 771

\bibitem[\protect\citeauthoryear{Hunter \& Lias}{Hunter \&
  Lias}{1998}]{Edward1998}
Hunter E. P.~L.,  Lias S.~G.,  1998, \mn@doi [Journal of Physical and Chemical
  Reference Data] {10.1063/1.556018}, 27, 413

\bibitem[\protect\citeauthoryear{Iqbal \& Wakelam}{Iqbal \&
  Wakelam}{2018}]{Iqbal2018}
Iqbal W.,  Wakelam V.,  2018, \mn@doi [Astronomy and Astrophysics]
  {10.1051/0004-6361/201732486}, 615, A20

\bibitem[\protect\citeauthoryear{Iza, Farenzena, Jalowy, Groeneveld  \& da
  Silveira}{Iza et~al.}{2006}]{Iza2006}
Iza P.,  Farenzena L.,  Jalowy T.,  Groeneveld K.,   da Silveira E.,  2006,
  \mn@doi [Nuclear Instruments and Methods in Physics Research Section B: Beam
  Interactions with Materials and Atoms]
  {https://doi.org/10.1016/j.nimb.2005.11.080}, 245, 61

\bibitem[\protect\citeauthoryear{Iza, Farenzena  \& da Silveira}{Iza
  et~al.}{2007}]{Iza2007}
Iza P.,  Farenzena L.,   da Silveira E.,  2007, \mn@doi [Nuclear Instruments
  and Methods in Physics Research Section B: Beam Interactions with Materials
  and Atoms] {https://doi.org/10.1016/j.nimb.2006.12.070}, 256, 483

\bibitem[\protect\citeauthoryear{Knoezinger, Beichert, Hermeling  \&
  Schrems}{Knoezinger et~al.}{1993}]{Knoezinger1993}
Knoezinger E.,  Beichert P.,  Hermeling J.,   Schrems O.,  1993, \mn@doi [The
  Journal of Physical Chemistry] {10.1021/j100109a013}, 97, 1324

\bibitem[\protect\citeauthoryear{Knyazheva, Khlebnikov, Kozulin, Kuzmina,
  Lyapin, Mutterer, Perkowski  \& Trzaska}{Knyazheva
  et~al.}{2006}]{Knyazheva2006}
Knyazheva G.,  Khlebnikov S.,  Kozulin E.,  Kuzmina T.,  Lyapin V.,  Mutterer
  M.,  Perkowski J.,   Trzaska W.,  2006, \mn@doi [Nuclear Instruments and
  Methods in Physics Research Section B: Beam Interactions with Materials and
  Atoms] {https://doi.org/10.1016/j.nimb.2006.04.071}, 248, 7

\bibitem[\protect\citeauthoryear{Lias}{Lias}{2001}]{LiasNIST}
Lias S.~G.,  2001, in Linstrom P.,  Mallard W.,  eds, , NIST Standard Reference
  Database Number 69.
National Institute of Standards and Technology, Gaithersburg MD, 20899, \url
  {http://webbook.nist.gov}

\bibitem[\protect\citeauthoryear{Macfarlane \& Torgerson}{Macfarlane \&
  Torgerson}{1976}]{Macfarlane1976}
Macfarlane R.,  Torgerson D.,  1976, \mn@doi [Science]
  {10.1126/science.1251202}, 191, 920

\bibitem[\protect\citeauthoryear{Maloteau, Mosley, Andre, Delhalle  \&
  Pickup}{Maloteau et~al.}{1992}]{Maloteau1992}
Maloteau M.,  Mosley D.~H.,  Andre J.~M.,  Delhalle J.,   Pickup B.~T.,  1992,
  \mn@doi [International Journal of Quantum Chemistry] {10.1002/qua.560440851},
  44, 563

\bibitem[\protect\citeauthoryear{Mart\'{\i}n-Dom\'enech, Mu\~noz Caro, Bueno
  \& Goesmann}{Mart\'{\i}n-Dom\'enech et~al.}{2014}]{Martin-Domenech2014}
Mart\'{\i}n-Dom\'enech R.,  Mu\~noz Caro G.~M.,  Bueno J.,   Goesmann F.,
  2014, \mn@doi [Astronomy and Astrophysics] {10.1051/0004-6361/201322824},
  564, A8

\bibitem[\protect\citeauthoryear{Martinez, Ponciano, Farenzena, Iza, Homem, de
  Brito, Wien  \& da Silveira}{Martinez et~al.}{2006}]{Martinez2006}
Martinez R.,  Ponciano C.,  Farenzena L.,  Iza P.,  Homem M.,  de Brito A.~N.,
  Wien K.,   da Silveira E.,  2006, \mn@doi [International Journal of Mass
  Spectrometry] {https://doi.org/10.1016/j.ijms.2006.04.005}, 253, 112

\bibitem[\protect\citeauthoryear{Martinez, Ponciano  \& da Silveira}{Martinez
  et~al.}{2012}]{Martinez2012}
Martinez R.,  Ponciano C.,   da Silveira E.,  2012, \mn@doi [The European
  Physical Journal D] {10.1140/epjd/e2012-30133-2}, 66, 251

\bibitem[\protect\citeauthoryear{Martinez, Bordalo, da Silveira  \&
  Boechat-Roberty}{Martinez et~al.}{2014}]{Martinez2014}
Martinez R.,  Bordalo V.,  da Silveira E.~F.,   Boechat-Roberty H.~M.,  2014,
  \mn@doi [Monthly Notices of the Royal Astronomical Society]
  {10.1093/mnras/stu1400}, 444, 3317

\bibitem[\protect\citeauthoryear{{Matthews} \& {Sears}}{{Matthews} \&
  {Sears}}{1983a}]{Matthews1983a}
{Matthews} H.~E.,  {Sears} T.~J.,  1983a, \mn@doi [The Astrophysical Journal]
  {10.1086/184001}, \href
  {https://ui.adsabs.harvard.edu/abs/1983ApJ...267L..53M} {267, L53}

\bibitem[\protect\citeauthoryear{{Matthews} \& {Sears}}{{Matthews} \&
  {Sears}}{1983b}]{Matthews1983b}
{Matthews} H.~E.,  {Sears} T.~J.,  1983b, \mn@doi [The Astrophysical Journal]
  {10.1086/161271}, \href
  {https://ui.adsabs.harvard.edu/abs/1983ApJ...272..149M} {272, 149}

\bibitem[\protect\citeauthoryear{McGuire, Burkhardt, Kalenskii, Shingledecker,
  Remijan, Herbst  \& McCarthy}{McGuire et~al.}{2018}]{McGuire2018}
McGuire B.~A.,  Burkhardt A.~M.,  Kalenskii S.,  Shingledecker C.~N.,  Remijan
  A.~J.,  Herbst E.,   McCarthy M.~C.,  2018, \mn@doi [Science]
  {10.1126/science.aao4890}, 359, 202

\bibitem[\protect\citeauthoryear{Mejías, Hamad  \& Lago}{Mejías
  et~al.}{2001}]{Mejias2001}
Mejías J.~A.,  Hamad S.,   Lago S.,  2001, \mn@doi [The Journal of Physical
  Chemistry B] {10.1021/jp003433x}, 105, 9872

\bibitem[\protect\citeauthoryear{Moore, Ferrante, Moore  \& Hudson}{Moore
  et~al.}{2010}]{Moore2010}
Moore M.~H.,  Ferrante R.~F.,  Moore W.~J.,   Hudson R.,  2010, \mn@doi [The
  Astrophysical Journal Supplement Series] {10.1088/0067-0049/191/1/96}, 191,
  96

\bibitem[\protect\citeauthoryear{Nastasi, Mayer  \& Hirvonen}{Nastasi
  et~al.}{1996}]{Nastasi1996}
Nastasi M.,  Mayer J.,   Hirvonen J.~K.,  1996, Ion stopping.
Cambridge University Press, p. 88–114, \mn@doi{10.1017/CBO9780511565007.006}

\bibitem[\protect\citeauthoryear{Pagani, Favre, Goldsmith, Bergin, Snell  \&
  Melnick}{Pagani et~al.}{2017}]{Pagani2017}
Pagani L.,  Favre C.,  Goldsmith P.~F.,  Bergin E.~A.,  Snell R.,   Melnick G.,
   2017, \mn@doi [Astronomy and Astrophysics] {10.1051/0004-6361/201730466},
  604, A32

\bibitem[\protect\citeauthoryear{Pankonin, Churchwell, Watson  \&
  Bieging}{Pankonin et~al.}{2001}]{Pankonin2001}
Pankonin V.,  Churchwell E.,  Watson C.,   Bieging J.~H.,  2001, \mn@doi [The
  Astrophysical Journal] {10.1086/322249}, 558, 194

\bibitem[\protect\citeauthoryear{Petrie, Freeman  \& McEwan}{Petrie
  et~al.}{1992}]{Petrie1992}
Petrie S.,  Freeman C.~G.,   McEwan M.~J.,  1992, \mn@doi [Monthly Notices of
  the Royal Astronomical Society] {10.1093/mnras/257.3.438}, 257, 438

\bibitem[\protect\citeauthoryear{Pino et~al.,}{Pino et~al.}{2019}]{Pino2019}
Pino T.,  et~al., 2019, \mn@doi [Astronomy and Astrophysics]
  {10.1051/0004-6361/201834855}, 623, A134

\bibitem[\protect\citeauthoryear{Pirim, Gann, McLain  \& Orlando}{Pirim
  et~al.}{2015}]{Pirim2015}
Pirim C.,  Gann R.,  McLain J.,   Orlando T.,  2015, \mn@doi [Icarus]
  {https://doi.org/10.1016/j.icarus.2015.06.006}, 258, 109

\bibitem[\protect\citeauthoryear{Plasil, Glos{\'{\i}}k  \& Zakouril}{Plasil
  et~al.}{1999}]{Plasil1999}
Plasil R.,  Glos{\'{\i}}k J.,   Zakouril P.,  1999, \mn@doi [Journal of Physics
  B: Atomic, Molecular and Optical Physics] {10.1088/0953-4075/32/14/322}, 32,
  3575

\bibitem[\protect\citeauthoryear{Ponciano, Martinez, Farenzena, Iza, da
  Silveira, Homem, de Brito  \& Wien}{Ponciano et~al.}{2006}]{Ponciano2006}
Ponciano C.,  Martinez R.,  Farenzena L.,  Iza P.,  da Silveira E.,  Homem M.,
  de Brito A.~N.,   Wien K.,  2006, \mn@doi [Journal of the American Society
  for Mass Spectrometry] {https://doi.org/10.1016/j.jasms.2006.04.018}, 17,
  1120

\bibitem[\protect\citeauthoryear{Rab, G\"udel, Padovani, Kamp, Thi, Woitke  \&
  Aresu}{Rab et~al.}{2017}]{Rab2017}
Rab C.,  G\"udel M.,  Padovani M.,  Kamp I.,  Thi W.-F.,  Woitke P.,   Aresu
  G.,  2017, \mn@doi [Astronomy and Astrophysics]
  {10.1051/0004-6361/201630241}, 603, A96

\bibitem[\protect\citeauthoryear{Ribeiro, Almeida, Garcia-Basabe, Wolff,
  Boechat-Roberty  \& Rocco}{Ribeiro et~al.}{2015}]{FAR2015}
Ribeiro F. d.~A.,  Almeida G.~C.,  Garcia-Basabe Y.,  Wolff W.,
  Boechat-Roberty H.~M.,   Rocco M. L.~M.,  2015, \mn@doi [Physical Chemistry
  Chemical Physics] {10.1039/C5CP05040E}, 17, 27473

\bibitem[\protect\citeauthoryear{Rong et~al.,}{Rong et~al.}{2015}]{Rong2015}
Rong J.,  et~al., 2015, \mn@doi [Monthly Notices of the Royal Astronomical
  Society] {10.1093/mnras/stv2406}, 455, 1428

\bibitem[\protect\citeauthoryear{Rosero, Hofner, Kurtz, Bieging  \&
  Araya}{Rosero et~al.}{2013}]{Rosero2013}
Rosero V.,  Hofner P.,  Kurtz S.,  Bieging J.,   Araya E.~D.,  2013, \mn@doi
  [The Astrophysical Journal Supplement Series] {10.1088/0067-0049/207/1/12},
  207, 12

\bibitem[\protect\citeauthoryear{Rothard, Domaracka, Boduch, Palumbo,
  Strazzulla, da Silveira  \& Dartois}{Rothard et~al.}{2017}]{Rothard2017}
Rothard H.,  Domaracka A.,  Boduch P.,  Palumbo M.~E.,  Strazzulla G.,  da
  Silveira E.~F.,   Dartois E.,  2017, \mn@doi [Journal of Physics B: Atomic,
  Molecular and Optical Physics] {10.1088/1361-6455/50/6/062001}, 50, 062001

\bibitem[\protect\citeauthoryear{Sailer, Pelc, Limão-Vieira, Mason, Limtrakul,
  Scheier, Probst  \& Märk}{Sailer et~al.}{2003}]{Sailer2003}
Sailer W.,  Pelc A.,  Limão-Vieira P.,  Mason N.,  Limtrakul J.,  Scheier P.,
  Probst M.,   Märk T.,  2003, \mn@doi [Chemical Physics Letters]
  {https://doi.org/10.1016/j.cplett.2003.09.118}, 381, 216

\bibitem[\protect\citeauthoryear{{Solomon}, {Jefferts}, {Penzias}  \&
  {Wilson}}{{Solomon} et~al.}{1971}]{Solomon1971}
{Solomon} P.~M.,  {Jefferts} K.~B.,  {Penzias} A.~A.,   {Wilson} R.~W.,  1971,
  \mn@doi [The Astrophysical Journal] {10.1086/180794}, \href
  {https://ui.adsabs.harvard.edu/abs/1971ApJ...168L.107S} {168, L107}

\bibitem[\protect\citeauthoryear{Stockdale, Davis, Compton  \& Klots}{Stockdale
  et~al.}{1974}]{Stockdale1974}
Stockdale J.~A.,  Davis F.~J.,  Compton R.~N.,   Klots C.~E.,  1974, \mn@doi
  [The Journal of Chemical Physics] {10.1063/1.1680900}, 60, 4279

\bibitem[\protect\citeauthoryear{Suess, Liu, Parthasarathy  \& Dunning}{Suess
  et~al.}{2003}]{Suess2003}
Suess L.,  Liu Y.,  Parthasarathy R.,   Dunning F.~B.,  2003, \mn@doi [The
  Journal of Chemical Physics] {10.1063/1.1628215}, 119, 12890

\bibitem[\protect\citeauthoryear{Sundqvist}{Sundqvist}{1993}]{Sundqvist1993}
Sundqvist B.~U.,  1993, \mn@doi [International Journal of Mass Spectrometry and
  Ion Processes] {https://doi.org/10.1016/0168-1176(93)80064-L}, 126, 1

\bibitem[\protect\citeauthoryear{Sundqvist \& Macfarlane}{Sundqvist \&
  Macfarlane}{1985}]{Sundqvist1985}
Sundqvist B.,  Macfarlane R.~D.,  1985, \mn@doi [Mass Spectrometry Reviews]
  {10.1002/mas.1280040403}, 4, 421

\bibitem[\protect\citeauthoryear{Taquet, L{\'{o}}pez-Sepulcre, Ceccarelli,
  Neri, Kahane  \& Charnley}{Taquet et~al.}{2015}]{Taquet2015}
Taquet V.,  L{\'{o}}pez-Sepulcre A.,  Ceccarelli C.,  Neri R.,  Kahane C.,
  Charnley S.~B.,  2015, \mn@doi [The Astrophysical Journal]
  {10.1088/0004-637x/804/2/81}, 804, 81

\bibitem[\protect\citeauthoryear{Thaddeus, Gottlieb, Gupta, Brünken, McCarthy,
  Ag{\'{u}}ndez, Gu{\'{e}}lin  \& Cernicharo}{Thaddeus
  et~al.}{2008}]{Thaddeus2008}
Thaddeus P.,  Gottlieb C.~A.,  Gupta H.,  Brünken S.,  McCarthy M.~C.,
  Ag{\'{u}}ndez M.,  Gu{\'{e}}lin M.,   Cernicharo J.,  2008, \mn@doi [The
  Astrophysical Journal] {10.1086/528947}, 677, 1132

\bibitem[\protect\citeauthoryear{Toumi, Couturier-Tamburelli, Chiavassa  \&
  Piétri}{Toumi et~al.}{2014}]{Toumi2014}
Toumi A.,  Couturier-Tamburelli I.,  Chiavassa T.,   Piétri N.,  2014, \mn@doi
  [The Journal of Physical Chemistry A] {10.1021/jp412481s}, 118, 2453

\bibitem[\protect\citeauthoryear{Toumi, Piétri, Chiavassa  \&
  Couturier-Tamburelli}{Toumi et~al.}{2016}]{Toumi2016}
Toumi A.,  Piétri N.,  Chiavassa T.,   Couturier-Tamburelli I.,  2016, \mn@doi
  [Icarus] {https://doi.org/10.1016/j.icarus.2014.10.042}, 270, 435

\bibitem[\protect\citeauthoryear{{Turner}, {Amano}  \& {Feldman}}{{Turner}
  et~al.}{1990}]{Turner1990}
{Turner} B.~E.,  {Amano} T.,   {Feldman} P.~A.,  1990, \mn@doi [The
  Astrophysical Journal] {10.1086/168321}, \href
  {https://ui.adsabs.harvard.edu/abs/1990ApJ...349..376T} {349, 376}

\bibitem[\protect\citeauthoryear{Vigren et~al.,}{Vigren
  et~al.}{2008}]{Vigren2008}
Vigren E.,  et~al., 2008, \mn@doi [Physical Chemistry Chemical Physics]
  {10.1039/B801566J}, 10, 4014

\bibitem[\protect\citeauthoryear{Vigren et~al.,}{Vigren
  et~al.}{2009}]{Vigren2009}
Vigren E.,  et~al., 2009, \mn@doi [The Astrophysical Journal]
  {10.1088/0004-637x/695/1/317}, 695, 317

\bibitem[\protect\citeauthoryear{Wakisaka, Abdoul-Carime, Yamamoto  \&
  Kiyozumi}{Wakisaka et~al.}{1998}]{Wakisaka1998}
Wakisaka A.,  Abdoul-Carime H.,  Yamamoto Y.,   Kiyozumi Y.,  1998, \mn@doi
  [Journal of the Chemical Society, Faraday Transactions] {10.1039/A705777F},
  94, 369

\bibitem[\protect\citeauthoryear{Walsh, Harada, Herbst  \& Millar}{Walsh
  et~al.}{2009}]{Catherine2009}
Walsh C.,  Harada N.,  Herbst E.,   Millar T.~J.,  2009, \mn@doi [The
  Astrophysical Journal] {10.1088/0004-637x/700/1/752}, 700, 752

\bibitem[\protect\citeauthoryear{Zeng et~al.,}{Zeng et~al.}{2018}]{Zeng2018}
Zeng S.,  et~al., 2018, \mn@doi [Monthly Notices of the Royal Astronomical
  Society] {10.1093/mnras/sty1174}, 478, 2962

\makeatother
\end{thebibliography}

\bsp	
\label{lastpage}
\end{document}